\documentclass[aps,prb,twocolumn,amsmath,amssymb,nofootinbib,
  superscriptaddress]{revtex4-2}

\usepackage{graphicx}
\usepackage{bm}
\usepackage{amsmath}
\usepackage{amssymb}
\usepackage{physics}
\usepackage[colorlinks=true,linkcolor=blue,urlcolor=blue,citecolor=blue]{hyperref}
\usepackage[usenames,svgnames]{xcolor}
\usepackage{makecell}
\usepackage[normalem]{ulem}
\usepackage{MnSymbol}
\usepackage{changes}
\usepackage{cancel}
\usepackage{soul}

\begin{document}

\newcommand{\Hop}{\hat{H}}
\newcommand{\Himp}{\hat{H}_{\rm imp}}
\newcommand{\Hint}{\hat{H}_{\rm int}}
\newcommand{\Hs}{\hat{H_{\rm S}}}
\newcommand{\Uimp}{\hat{U}_{\rm imp}}
\newcommand{\gHimp}{\mathcal{H}_{\rm imp}}
\newcommand{\gUimp}{\mathcal{U}_{\rm imp}}
\newcommand{\Hbath}{\hat{H}_{\rm bath}}
\newcommand{\Hhyb}{\hat{H}_{\rm hyb}}
\newcommand{\scgap}{\Delta^{\rm sc}}

\newcommand{\sop}{\hat{s}}
\newcommand{\bop}{\hat{b}}
\newcommand{\bdop}{\hat{b}^{\dagger}}
\newcommand{\aop}{\hat{a}}
\newcommand{\adop}{\hat{a}^{\dagger}}
\newcommand{\sgp}{\hat{\sigma}^+}
\newcommand{\sgx}{\hat{\sigma}^x}
\newcommand{\sgy}{\hat{\sigma}^y}
\newcommand{\sgz}{\sigma_z}
\newcommand{\nop}{\hat{n}}

\newcommand{\Topc}{\mathcal{T}_{\mathcal{C}}}

\newcommand{\cop}{\hat{c}}
\newcommand{\cdop}{\hat{c}^{\dagger}}
\newcommand{\Cop}{\hat{C}}
\newcommand{\Cdop}{\hat{C}^{\dagger}}
\newcommand{\hc}{{\rm H.c.}}
\newcommand{\rhotot}{\hat{\rho}_{\mathrm{tot}}}
\newcommand{\rhoop}{\hat{\rho}}
\newcommand{\rhoimp}{\hat{\rho}_{\mathrm{imp}}}
\newcommand{\rhobath}{\hat{\rho}_{\mathrm{bath}}^{\rm th}}
\newcommand{\Zimp}{Z_{{\rm imp}}}
\newcommand{\Zbath}{Z_{\mathrm{bath}}}
\newcommand{\mea}{\mathcal{D}}
\newcommand{\gK}{\mathcal{K}}
\newcommand{\gI}{\mathcal{I}}
\newcommand{\gF}{\mathcal{F}}
\newcommand{\bolda}{\bm{a}}
\newcommand{\boldabar}{\bar{\bm{a}}}
\newcommand{\abar}{\bar{a}}
\newcommand{\im}{{\rm i}}
\newcommand{\contour}{\mathcal{C}}
\newcommand{\gA}{\mathcal{A}}
\newcommand{\gB}{\mathcal{B}}
\newcommand{\gM}{\mathcal{M}}
\newcommand{\boldeta}{\bm{\eta}}
\newcommand{\boldetabar}{\bar{\bm{\eta}}}
\newcommand{\etabar}{\bar{\eta}}
\newcommand{\parity}{\mathcal{P}}
\newcommand{\current}{\mathcal{J}}
\newcommand{\JW}{{\rm JW}}
\newcommand{\pronyerror}{\varsigma_p}
\newcommand{\WI}{{\rm W}^I}
\newcommand{\WII}{{\rm W}^{II}}
\newcommand{\branch}{\mathcal{C}}

\newcommand{\Aop}{\hat{\alpha}}
\newcommand{\Adop}{\hat{\alpha}^{\dagger}}
\newcommand{\Gop}{\mathcal{G}}
\newcommand{\boldGop}{\bm{\mathcal{G}}}

\newcommand{\Xop}{\hat{X}}
\newcommand{\Xdop}{\hat{X}^{\dagger}}
\newcommand{\Xbar}{\bar{X}}
\newcommand{\boldXbar}{\bm{\bar{X}}}
\newcommand{\boldX}{\bm{X}}
\newcommand{\Yop}{\hat{Y}}
\newcommand{\Ydop}{\hat{Y}^{\dagger}}
\newcommand{\Ybar}{\bar{Y}}
\newcommand{\boldYbar}{\bm{\bar{Y}}}
\newcommand{\boldY}{\bm{Y}}
\newcommand{\boldDelta}{\bm{\Delta}}

\newcommand{\boldpsi}{\bm{\psi}}
\newcommand{\boldpsibar}{\bar{\bm{\psi}}}

\newcommand{\EqDef}{\stackrel{\mathrm{def}}{=}}

\newcommand{\gcc}[1]{{\color{red}#1}}
\definechangesauthor[name={RF},color=green]{RF}

\newcommand{\snu}{College of Physics and Electronic Engineering, and Center for Computational Sciences, Sichuan Normal University, Chengdu 610068, China}

\newcommand{\nudt}{College of Science, National University of Defense Technology, Changsha 410073, China}

\newcommand{\jianga}{College of Advanced Interdisciplinary Studies, National University of Defense Technology, Changsha, China}
\newcommand{\jiangb}{Hunan Research Center of the Basic Discipline for Physical States, Changsha, China}

\newcommand{\chena}{Hunan Key Laboratory of Mechanism and technology of Quantum Information, Changsha, China}

\newcommand{\yuecm}{State Key Laboratory of Quantum Functional Materials, Department of Physics, and Guangdong Basic Research Center of Excellence for Quantum Science, Southern University of Science and Technology (SUSTech), Shenzhen 518055, China}

\title{Grassmann time-evolving matrix product operators for superconducting quantum impurity problem }

\author{Chu Guo}
\thanks{These authors contribute equally to this work}
\affiliation{\nudt}
\affiliation{\chena}

\author{Wei Wu}
\thanks{These authors contribute equally to this work}
\affiliation{\nudt}
\affiliation{\chena}

\author{Xiansong Xu}
\affiliation{\snu}

\author{Ping-Xing Chen}
\affiliation{\nudt}
\affiliation{\chena}

\author{Changming Yue}
\email{yuecm@sustech.edu.cn}
\affiliation{\yuecm}

\author{Tian Jiang}
\email{tjiang@nudt.edu.cn}
\affiliation{\nudt}
\affiliation{\jianga}
\affiliation{\jiangb}

\author{Ruofan Chen}
\email{physcrf@sicnu.edu.cn}
\affiliation{\snu}

\date{\today}


\begin{abstract}
The Grassmann time-evolving matrix product operator (GTEMPO) method, which represents the Feynman-Vernon influence functional as a temporal matrix product state, has been shown to be a flexible and potentially scalable solution for fermionic quantum impurity problems. 
In this work, we extend GTEMPO to solve fermionic impurity problems in the Nambu formalism, in which the impurity is coupled to a superconducting bath. A key insight is that by employing the Bogoliubov transformation for the superconducting bath, one could obtain the analytic expression of the Feynman-Vernon influence functional in a similar form to the case of a normal bath, after which the core algorithms of GTEMPO can be straightforwardly adapted. 
We demonstrate the accuracy of our method by benchmarking it against exact diagonalization in several exactly solvable cases, 
and against the continuous-time quantum Monte Carlo method using converged dynamical mean field theory (DMFT) iterations on the imaginary contour 
in the non-integrable case.
In all cases, we perform both imaginary- and real-time calculations to illustrate the flexibility of our method.
These results illustrate that our method, termed Nambu-GTEMPO, could be potentially useful as a quantum impurity solver to study the superconducting states in both equilibrium and non-equilibrium systems.
\end{abstract}

\maketitle

\section{Introduction}
The problem of an impurity coupled to structured baths describes a ubiquitous phenomenon in condensed matter physics, quantum chemistry, and materials science~\cite{anderson1961-localized,Georges2016,mahan2000-many,weiss1993-quantum,MetznerVollhardt1989,GeorgesKotliar1992,GeorgesRozenberg1996}. Among various bath configurations, the superconducting bath holds particular significance due to its relevance in understanding Andreev reflection~\cite{BlonderKlapwijk1982,CuevasYeyati1996}, Yu-Shiba-Rusinov states~\cite{AndersenPaaske2011}, Majorana zero modes~\cite{Kitaev2001,Alicea2012}, and the interplay between local correlations and superconducting order~\cite{Sakurai1970,HartmannZittartz1971,MatsuuraTamifusa1977,BalatskyZhu2006,YazdaniEigler1997,FrankePascual2011}. A prototypical theoretical framework to study such systems is the quantum impurity problem (QIP) which considers a localized impurity coupled to a superconducting fermionic bath. QIP with a superconducting bath is also the central problem within the framework of dynamical mean-field theory (DMFT) in the Nambu–Gor'kov formalism~\cite{Gorkov1958,Nambu1960}. 
Nambu-DMFT provides a non-perturbative framework that captures the dynamical nature of electronic correlations and pairing, including the frequency-dependent anomalous self-energy and the associated pairing glue. Its combination with cluster extensions further enables the inclusion of short-range spatial correlations, which are essential for describing unconventional superconductivity beyond the mean-field and weak coupling approaches.
Over the past two decades, such methods have been extensively applied to investigate the interplay between correlation and pairing mechanism in the strongly correlated electron systems, including the Hubbard models~\cite{Lichtenstein2000,CaffarelKrauth1994,Kancharla2008,Civelli2009,Sakai2016PRL,Sakai2016PRB,Sakai2018,Foley2019,Sordi2012,GullMillis2013,DongGullMillis2022,YueWerner2022,LiuWu2025CPL}, alkali-doped fullerides~\cite{Capone2002,Capone2009,Nomura2015SciAdv,Yue2021Fulleride,Yue2023Cs3C60,Yue2022PRL}, cuprates~\cite{Sakai2016PRL,Sakai2018,Sentef2011,PhysRevX.15.021071,SakaiImada2026PRX}, nickelates~\cite{RQHePRB2024,MoWu2025,LiuWu2025CPL,Nomura2025PRB}, FeSe~\cite{Yue2021FeSe}, and even quasicrystalline systems~\cite{SakaiTakemori2022,Fukushima2023}.
However, although being formally simple and elegant, solving these QIPs under general parameter settings still remains numerically challenging, 
and there is a longstanding research interest to develop efficient and accurate numerical methods for such problems.


A number of numerical methods have been applied to solve QIPs in presence of a superconducting bath, including continuous-time quantum Monte Carlo (CTQMC)~\cite{LuitzAssaad2010,LuitzMeden2012,Yue2021FeSe,YueWerner2022}, numerical renormalization group~\cite{SatoriShimizu1992,SakaiSatori1993,BauerHewson2007,KarraschMeden2008,RokAguado2015}, exact diagonalization (ED)~\cite{ZhangLu2026}, time-evolving matrix product state (MPS)~\cite{WautersBurrello2023}. Among these methods, the CTQMC methods are known to be numerically exact, but are only efficient on the imaginary-time axis~\cite{RubtsovLichtenstein2004,RubtsovLichtenstein2005,WernerMillis2006,GullTroyer2008,ChanMillis2009,GullWerner2011}. The rest methods can directly work on the real-time axis, but they generally suffer from uncontrolled errors, such as the bath discretization error~\cite{InesWolf2015}, and could be complicated for finite-temperature baths~\cite{KohnSantoro2021,KohnSantoro2022}.

The GTEMPO method, proposed by some of the authors, has illustrated itself to be a flexible and accurate method for QIPs in recent years. 
GTEMPO is rooted from the quasi-adiabatic propagator path integral (QuAPI)~\cite{makarov1994-path,makri1995-numerical} and the time-evolving matrix product operator (TEMPO)~\cite{StrathearnLovett2018} methods, where the latter two methods are used for bosonic quantum impurity problems. QuAPI makes use of the Feynman-Vernon influence functional (IF) to represent the integrand of the discretized impurity path integral (IPI) as a tensor on the time axis, which is referred to as the augmented density tensor (ADT), while TEMPO further represents this tensor as a temporal MPS.
GTEMPO represents the ADT as a Grassmann MPS (GMPS), as the ADT in the fermionic case is naturally a Grassmann tensor~\cite{gu2013-efficient,yoshimura2018-calculation,akiyama2021-more}.
Similar to TEMPO, GTEMPO has only two error sources: (i) the Trotter decomposition error of the IPI and (ii) the MPS bond truncation error, which can both be well-controlled. Since GTEMPO only relies on the analytical expression of the IPI (essentially the IF), it can be straightforwardly applied on the imaginary~\cite{ChenGuo2024b}, real (Keldysh)~\cite{ChenGuo2024a} and the L-shaped Kadanoff~\cite{ChenGuo2024g} contours, and can consider the simultaneous presence of bosonic and fermionic baths as the influence of a bosonic bath can be treated as retarded interaction in the IPI~\cite{ChenGuo2025}. In addition, GTEMPO is potentially scalable to large impurity problems by iteratively tracing out the parts of impurity that are not directly used in the interested observables~\cite{SunGuo2025b}.


However, currently only normal fermionic baths are supported in GTEMPO. A major obstacle to apply GTEMPO for superconducting baths is that the analytic expression of the IF is currently not available. This gap is filled in this work. We first derive the Feynman-Vernon IF in this case, by employing the Bogoliubov transformation to transform the superconducting bath into the same form of a normal bath.
The obtained IF has a similar quadratic form to the standard case, therefore we can straightforwardly adapt the core algorithms of the standard GTEMPO to this case. 
In our numerical simulations, we apply our method to study the superconducting Anderson impurity model (SAIM) which describes a localized electron (with nonzero interaction) that is coupled to a superconducting bath. We perform converged DMFT calculations on the imaginary-time axis and benchmark our results against the CTQMC results, where we show a step-by-step match between these two methods. We also study the non-equilibrium real-time evolution of the SAIM from an uncorrelated impurity-bath initial state, which is hard for non-equilibrium CTQMC suffering from the dynamical sign problem~\cite{TroyerWiese2005,Eckstein2010,CohenPRL2015}. These results illustrate that our method is an accurate and flexible solver for QIPs with a superconducting bath, which could be useful in DMFT~\cite{GullWerner2011} and non-equilibrium DMFT~\cite{AokiWerner2014} in the Nambu formalism.

This paper is organized as follows. In Sec.~\ref{sec:model}, we present the general Hamiltonian for the QIPs considered in this work. In Sec.~\ref{sec:gtempo}, we briefly review the GTEMPO method for normal fermionic baths. In Sec.~\ref{sec:method}, we present the details of our Nambu-GTEMPO method for superconducting baths. 
In Sec.~\ref{sec:results} we show our numerical results for the SAIM. We conclude in Sec.~\ref{sec:summary}.


\section{Model description}\label{sec:model}
The total Hamiltonian of a QIP can be written as
\begin{align}
\Hop = \Himp + \Hint = \Himp + \Hhyb + \Hbath,
\end{align}
where $\Himp$ and $\Hbath$ denote the impurity and bath Hamiltonians respectively, and $\Hhyb$ denotes the coupling between the impurity and bath. In the second equality we have defined $\Hint=\Hhyb+\Hbath$, which encodes all the bath effects. For QIPs, $\Hbath$ is generally assumed to be non-interacting. Another commonly used assumption is that $\Hhyb$ is linear in terms of the bath creation and annihilation operators, which, together with the previous assumption, constitute the basic requirements to derive the analytic expression of the Feynman-Vernon IF~\cite{FeynmanVernon1963}.

In this work, we will focus on the SAIM, although our method is completely general for any fermionic QIPs satisfying the above two assumptions.
The Hamiltonian of the SAIM can be written as
\begin{align}
&\Himp = \varepsilon_d\sum_{\sigma=\uparrow\downarrow} \adop_{\sigma}\aop_{\sigma} + U\adop_{\uparrow}\adop_{\downarrow}\aop_{\downarrow}\aop_{\uparrow}, \label{eq:Himp} \\
&\Hhyb = \sum_{k,\sigma} V_k(\adop_{\sigma}\cop_{k, \sigma} + \cdop_{k,\sigma}\aop_{\sigma}), \label{eq:Hhyb} \\
&\Hbath = \sum_{k, \sigma} \varepsilon_{k, \sigma}\cdop_{k, \sigma}\cop_{k, \sigma} + \sum_{k}(\scgap_k\cdop_{k,\uparrow}\cdop_{-k,\downarrow} + \hc). \label{eq:Hbath}
\end{align}
Here $\adop_{\sigma}$, $\aop_{\sigma}$ and $\cdop_{k, \sigma}$, $\cop_{k, \sigma}$ are the fermionic creation and annihilation operators of the impurity and the bath respectively, $\varepsilon_d$ and $U$ are the on-site energy and interaction strength of the localized electron orbital, $\varepsilon_{k,\sigma}$ is the band energy, $V_k$ is the coupling strength between the impurity and bath. We have considered the Bardeen-Cooper-Schrieffer (BCS) Hamiltonian with s-wave pairing~\cite{BCS1957} in Eq.(\ref{eq:Hbath}) for the superconducting bath, where $\scgap_k$ denotes the energy gap. For real-time evolution, we will consider an uncorrelated impurity-bath initial state
\begin{align}\label{eq:rho0}
\rhoop_0 = \rhoimp \otimes \rhobath,
\end{align}
where $\rhoimp$ could be some arbitrary impurity state, $\rhobath = \exp(-\beta \Hbath)$ is the (unnormalized) thermal state of the bath with inverse temperature $\beta$.

\section{Review of GTEMPO for the Anderson impurity model}\label{sec:gtempo}
In this section, we will briefly review the GTEMPO method for the standard Anderson impurity model (AIM) in which the impurity is coupled to a normal fermionic bath, as most steps of it can be directly reused in our extension for superconducting baths. We will keep our discussions general for any contours unless it is necessary to refer to some specific contour.

\subsection{The impurity path integral}
For the AIM, one can trace out the bath degrees of freedom in the total path integral of the problem, and obtain a reduced path integral for the impurity degrees of freedom only, i.e., the IPI. In terms of the Grassmann trajectories $\abar_{\sigma}(\tau)$ and $a_{\sigma}(\tau)$, where $\tau$ is a specific time on the contour that the problem is defined, $\abar_{\sigma}(\tau)$ and $a_{\sigma}(\tau)$ are a conjugate pair of Grassmann variables (GVs) at time $\tau$, one could write the impurity partition function, denoted as $\Zimp$, as an IPI~\cite{kamenev2009-keldysh,negele1998-quantum,kadanoff1962-quantum}:
\begin{align}
\Zimp= \int\mea[\boldabar \bolda] \gK[\boldabar \bolda]\gI[\boldabar \bolda],
\end{align}
where $\boldabar_{\sigma} = \{\abar_{\sigma}(\tau)\}$, $\bolda_{\sigma}=\{a_{\sigma}(\tau)\}$, $\boldabar=\{\boldabar_{\uparrow}, \boldabar_{\downarrow}\}$, $\bolda=\{\bolda_{\uparrow}, \bolda_{\downarrow}\}$ for briefness. The measure $\mea[\boldabar \bolda] $ for the Grassmann trajectories is defined as
\begin{align}
\mea[\boldabar \bolda] = \prod_{\sigma, \tau}\dd\abar_{\sigma}(\tau)\dd a_{\sigma}(\tau).
\end{align}
$\gK[\boldabar \bolda]$ denotes the contribution from the bare impurity Hamiltonian $\Himp$, which can be formally written as
\begin{align}\label{eq:gK}
\gK[\boldabar \bolda] = \exp\left[-\im \int_{\contour}\gHimp(\tau)\right],
\end{align}
where $\gHimp(\tau )$ is obtained from $\Himp(\tau)$ by making the substitutions $\aop_{\sigma}(\tau)\rightarrow a_{\sigma}(\tau)$ and $\adop_{\sigma}(\tau)\rightarrow \abar_{\sigma}(\tau)$, $\contour$ denotes the contour on which $\Zimp$ is defined, which can be the imaginary contour ($0\rightarrow\beta$), the Keldysh contour ($0\rightarrow t\rightarrow 0$) or the L-shaped Kadanoff contour ($0\rightarrow t\rightarrow 0\rightarrow\beta$). $\gI[\boldabar \bolda]$ denotes the IF which encodes all the bath influence, determined by $\Hint$ (and $\rhoop_0$) only. For the standard AIM, the two spin species are decoupled in $\Hint$, and $\gI[\boldabar \bolda]$ can be further decomposed into
\begin{align}\label{eq:IFdecompose}
\gI[\boldabar \bolda]=\prod_{\sigma}\gI_{\sigma}[\boldabar_{\sigma} \bolda_{\sigma}],
\end{align}
where $\gI_{\sigma}[\boldabar_{\sigma} \bolda_{\sigma}]$ denotes the IF for each spin species and can be written as
\begin{align}\label{eq:IFnormal}
\gI_{\sigma}[\boldabar_{\sigma} \bolda_{\sigma}] =\exp\left[-\int_{\contour}\dd\tau\int_{\contour}\dd\tau'\abar_{\sigma}(\tau)\Delta(\tau, \tau')a_{\sigma}(\tau') \right].
\end{align}
Here $\Delta(\tau,\tau')$ is the hybridization function which can be calculated as~\cite{chen2025-path}
\begin{align}\label{eq:hybridization}
\Delta(\tau,\tau') = \im \int d\varepsilon J(\varepsilon)D_{\varepsilon}(\tau, \tau'),
\end{align}
with $J(\varepsilon)=\sum_k V_k^2\delta(\varepsilon-\varepsilon_k)$ the bath spectral function (BSF), and $D_{\varepsilon}(\tau, \tau')=-\im\langle \Topc \cop_{\varepsilon}(\tau)\cdop_{\varepsilon}(\tau')\rangle_{\rm bath} $ the free-bath contour-ordered Green's function ($\Topc$ is the contour-ordering operator which orders the operators with smaller times on the contour to the right, $\langle\cdots\rangle_{\rm bath}$ means the expectation value over $\rhobath$). From Eq.(\ref{eq:IFnormal}), we can see that the hybridization function essentially encodes all the influence of the bath. In addition, we point out that the exponent in Eq.(\ref{eq:IFnormal}), i.e., $\abar_{\sigma}(\tau)\Delta(\tau, \tau')a_{\sigma}(\tau')$, can be formally related to $\Hhyb$ in Eq.(\ref{eq:Hhyb}) as follows: the two GVs $\abar_{\sigma}$ and $a_{\sigma}$ can be obtained from the two impurity operators $\adop_{\sigma}$ and $\aop_{\sigma}$, which are coupled to the bath operators $\cop_{k,\sigma}$ and $\cdop_{k, \sigma}$ respectively, via the mapping $\adop_{\sigma}\rightarrow \abar_{\sigma}, \aop_{\sigma}\rightarrow a_{\sigma}$. This observation is crucial for a formal understanding of our derivation of the Feynman-Vernon IF for the SAIM later.

\subsection{The GTEMPO method}

Similar to TEMPO, the first step of GTEMPO is to discretize the continuous Grassmann trajectories $\abar_{\sigma}(\tau)$ and $a_{\sigma}(\tau)$ into discrete sets of GVs, denoted as $\abar_{\sigma,j_{\zeta}}=\abar_{\sigma}(j_{\zeta}\delta\tau)$ and $a_{\sigma,j_{\zeta}}=a_{\sigma}(j_{\zeta}\delta\tau)$ where $\delta\tau$ denotes the discrete time step size, $j_{\zeta}$ denotes the $j$th time step on the $\zeta$th branch and we will use $\zeta=o,+, -$ for the imaginary ($0\rightarrow \beta$), forward ($0\rightarrow t$) and backward ($t\rightarrow 0$) branches respectively. The discrete $\gK[\boldabar\bolda]$ can be obtained from Eq.(\ref{eq:gK}) by inserting complete coherent-state basis set at each discrete time step. For example, on the imaginary contour, it can be written as
\begin{align}\label{eq:gKdis} 
\gK[\boldabar\bolda] = \langle -\bolda_{,1_o}\vert \Uimp^{o}\vert \bolda_{,M_o}\rangle \cdots  \langle \bolda_{,2_o}\vert \Uimp^{o}\vert \bolda_{,1_o}\rangle, 
\end{align}
where $\Uimp^o = \exp(-\delta \tau\Himp)$ denotes the discrete bare impurity propagator on the imaginary branch, with imaginary time step size $\delta\tau$, and we have used $\bolda_{, j_o} = \{a_{\uparrow, j_o}, a_{\downarrow, j_o}\}$. The discrete expressions of $\gK[\boldabar\bolda]$ on the Keldysh and Kadanoff contours could be found in Ref.~\cite{ChenGuo2024a} and Ref.~\cite{ChenGuo2024g} respectively, which slightly differ from Eq.(\ref{eq:gKdis}) due to the different boundary conditions on these contours.
The discrete $\gI_{\sigma}[\boldabar_{\sigma}\bolda_{\sigma}]$ corresponding to Eq.(\ref{eq:IFnormal}) can be written in a unified form that is valid for all contours:
\begin{align}\label{eq:gInormaldis}
&\gI_{\sigma}[\boldabar_{\sigma}\bolda_{\sigma}] = \exp\left(-\sum_{i_{\zeta},j_{\zeta'}} \abar_{\sigma, i_{\zeta}} \Delta_{i_{\zeta} j_{\zeta'}}a_{\sigma, j_{\zeta'}} \right),
\end{align}
where the discretized hybridization function $\Delta_{i_{\zeta}j_{\zeta'}}$ can be obtained using the QuAPI scheme~\cite{makarov1994-path,makri1995-numerical,ChenGuo2024a}
\begin{align}\label{eq:quapi}
\Delta_{i_{\zeta}j_{\zeta'}} = \int_{i_{\zeta}\delta\tau}^{(i_{\zeta}+1)\delta\tau}\dd\tau \int_{j_{\zeta'}\delta\tau}^{(j_{\zeta'}+1)\delta\tau}\dd\tau'\Delta(\tau, \tau').
\end{align}
By choosing a specfic ordering for all the GVs, one could systematically write the discrete $\gK$ and $\gI_{\sigma}$ as GMPSs (in practice, the orderings used in Ref.~\cite{ChenGuo2024b}, Ref.~\cite{ChenGuo2024b} and Ref.~\cite{ChenGuo2024g} could be good choices for the three contours respectively). 
For the AIM, $\gK$ can be built as a GMPS exactly~\cite{ChenGuo2024g}. $\gI_{\sigma}$ can be built as a GMPS using the \textit{partial IF} algorithm, for example. The partial IF algorithm explores the fact that each term on the exponent of Eq.(\ref{eq:gInormaldis}) commute with each other due to the Grassmann algebra, as a result, one could rewrite it as
\begin{align}
\gI_{\sigma}[\boldabar_{\sigma}\bolda_{\sigma}] = \prod_{i_{\zeta}}\exp\left(-\abar_{\sigma, i_{\zeta}}\sum_{j_{\zeta'}}\Delta_{i_{\zeta}j_{\zeta'}}a_{\sigma, j_{\zeta'}}\right),
\end{align}
where each term in the product is referred to as a partial IF. Crucially, each partial IF can be exactly written as a GMPS with bond dimension $2$ only~\cite{GuoChen2024d}, as all the terms on its exponent share the same GV $\abar_{\sigma, i_{\zeta}}$. As a result, the discrete $\gI_{\sigma}$ can be built as a GMPS using $M$ GMPS multiplications where $M$ is the number of all possible $i_{\zeta}$ (the multiplication of two GMPSs originates from the product of two Grassmann tensors~\cite{GuoChen2024d}, which is similar to the \textit{element-wise product} between two normal tensors used in QuAPI and TEMPO~\cite{GuoChen2026a}). Once $\gK$ and $\gI_{\sigma}$ are built as GMPSs, one could obtain the ADT as a GMPS by multiplying them together:
\begin{align}\label{eq:gAnormal}
\gA[\boldabar \bolda] = \gK[\boldabar \bolda] \gI[\boldabar \bolda]=\gK[\boldabar \bolda]\prod_{\sigma} \gI[\boldabar_{\sigma} \bolda_{\sigma}],
\end{align}
where we have used Eq.(\ref{eq:IFdecompose}) in the second equality.
The ADT contains all the information of the impurity dynamics, with which one can easily calculate any observables of the impurity. 
In practice, the multiplication in Eq.(\ref{eq:gAnormal}) does not need to be actually performed, but only needs to be calculated on the fly when calculating observables~\cite{ChenGuo2024a}.

To summarize, the GTEMPO method works in three steps: (1) discretizing the Grassmann trajectories and obtaining the discrete expressions of $\gK$ and $\gI_{\sigma}$ as in Eq.(\ref{eq:gKdis}) and Eq.(\ref{eq:gInormaldis}); (2) constructing the discrete $\gK$ and $\gI_{\sigma}$ as GMPSs, and obtaining the ADT by multiplying them together (on the fly) using Eq.(\ref{eq:gAnormal}); (3) calculating impurity observables based on $\gA$.


\section{The Nambu-GTEMPO method}\label{sec:method}
The standard AIM can be viewed as a special instance of the SAIM with $\scgap_k = 0$. As $\scgap_k$ only affects $\Hbath$, the only change we need to make on top of the GTEMPO method, is to change the way to build the IF as a GMPS, that is, only Eq.(\ref{eq:IFdecompose}) and Eq.(\ref{eq:IFnormal}) need to be changed.

In the following, we will first derive the analytic expression of the Feynman-Vernon IF for the SAIM, in correspondence with Eq.(\ref{eq:IFnormal}) for the AIM, and then we adapt the partial IF algorithm used in GTEMPO to build this new IF as a GMPS.

\subsection{The Feynman-Vernon IF for the SAIM}

For nonzero $\scgap_k$, the two spin species are mixed in $\Hbath$, as a result $\gI[\boldabar\bolda]$ can no longer be decomposed into the product of the two independent IFs $\gI_{\uparrow}$ and $\gI_{\downarrow}$ for each spin species.
Nevertheless, by employing the Bogoliubov transformation for the BCS bath:
\begin{align}
    &\Aop_{k,\uparrow}=u_k\cop_{k,\uparrow}-v_k\cdop_{-k,\downarrow},\quad \Adop_{k,\uparrow}=u\cdop_{k,\uparrow}-v_k\cop_{-k,\downarrow},\\
    &\Aop_{-k,\downarrow}=u_k\cop_{-k,\downarrow}+v_k\cdop_{k,\uparrow},\quad \Adop_{-k,\downarrow}=u_k\cdop_{-k,\downarrow}+v_k\cop_{k,\uparrow},
\end{align}
where $\Adop_{k, \sigma}$ and $\Aop_{k,\sigma}$ are the fermionic creation and annihilation operators in the new representation, we can convert $\Hbath$in Eq.(\ref{eq:Hbath}) into the form of a normal bath:
\begin{align}
  \label{eq:Hbathn}
  \Hbath'=E_0+\sum_k\xi_k(\Adop_{k,\uparrow}\Aop_{k,\uparrow}+\Adop_{-k,\downarrow}\Aop_{-k,\downarrow}),
\end{align}
where $E_0$ is a global constant term which does not affect the impurity observables and will thus be neglected, and
\begin{align}\label{eq:dispersion}
  \xi_k=\sqrt{\varepsilon_k^2+\abs{\scgap_k}^2},
\end{align}
is the dispersion relation in the Bogoliubov representation.
The complex numbers $u_k$ and $v_k$ in the Bogoliubov transformation are defined as
\begin{align}\label{eq:uv}
  u_k= \sqrt{\frac{1}{2}\qty(1+\frac{\varepsilon_k}{\xi_k})},\quad
  v_k=\frac{\scgap_k}{|\scgap_k|}\sqrt{\frac{1}{2}\qty(1-\frac{\varepsilon_k}{\xi_k})},
\end{align}
which satisfy $\abs{u_k}^2+\abs{v_k}^2=1$.
After the transformation, the hybridization Hamiltonian $\Hhyb$ in Eq.(\ref{eq:Hhyb}) becomes
\begin{align}
    \label{eq:Hhybn}
    \Hhyb' =&\sum_kV_k\left[(u_k^*\adop_{\uparrow}+v_k^*\aop_{\downarrow})\Aop_{k,\uparrow}
    +\Adop_{k,\uparrow}(u_k\aop_{\uparrow}+v_k\adop_{\downarrow}) \right. \nonumber \\
    &\left. +(u_k^*\adop_{\downarrow}-v_k^*\aop_{\uparrow})\Aop_{-k,\downarrow}
    +\Adop_{-k,\downarrow}(u_k\aop_{\downarrow}-v_k\adop_{\uparrow})\right].
\end{align}
To draw closer connection to Eq.(\ref{eq:Hhyb}), we define $\Xdop_k = u_k^*\adop_{\uparrow}+v_k^*\aop_{\downarrow}$ and $\Ydop_{-k} = u_k^*\adop_{\downarrow}-v_k^*\aop_{\uparrow}$, 
then Eq.(\ref{eq:Hhybn}) becomes
\begin{align}
    \label{eq:Hhybn2}
    \Hhyb' =\sum_kV_k\left(\Xdop_k\Aop_{k,\uparrow}
    +\Ydop_{-k}\Aop_{-k,\downarrow} + \hc\right).
\end{align}
Using the substitutions $\adop_{\uparrow}\rightarrow\Xdop_k$ and $\adop_{\downarrow}\rightarrow\Ydop_{-k}$, we can see that Eq.(\ref{eq:Hhybn2}) has exactly the same form as Eq.(\ref{eq:Hhyb}). The major difference is that $\adop_{\uparrow}$ and $\adop_{\downarrow}$ in Eq.(\ref{eq:Hhyb}) are ``local'' impurity operators acting on different spin species independently, while $\Xdop_k$ and $\Ydop_{-k}$ are ``global'' impurity operators simultaneously acting on both spin species. Nevertheless, since $\Hint' = \Hhyb' + \Hbath'$ does not mix the two spin species of the new bath modes (instead, the two spin species of the impurity are mixed in the new representation), we can still integrate out each spin species of the bath independently and obtain~\cite{chen2025-path}
\begin{align}\label{eq:IFdecomposen}
\gI[\boldabar\bolda] = \gI_{X}[\boldXbar\boldX]\gI_{Y}[\boldYbar \boldY],
\end{align}
in correspondence with Eq.(\ref{eq:IFdecompose}). The two terms on the right hand side of Eq.(\ref{eq:IFdecomposen}) can be explicitly written as 
\begin{align}
&\gI_{X}[\boldXbar \boldX] =\exp\left[-\int_{\contour}\dd\tau\int_{\contour}\dd\tau'\Xbar_k(\tau)\Delta_k(\tau, \tau')X_k(\tau') \right]; \label{eq:gX} \\
&\gI_{Y}[\boldYbar \boldY] =\exp\left[-\int_{\contour}\dd\tau\int_{\contour}\dd\tau'\Ybar_{-k}(\tau)\Delta_{-k}(\tau, \tau')Y_{-k}(\tau') \right], \label{eq:gY}
\end{align}
where we have used
\begin{align}
&\Xbar_k = u_k^*\abar_{\uparrow}+v_k^*a_{\downarrow}, \quad X_k = u_ka_{\uparrow}+v_k\abar_{\downarrow}; \label{eq:XX} \\
&\Ybar_{-k} = u_k^*\abar_{\downarrow}-v_k^*a_{\uparrow}, \quad Y_{-k} = u_ka_{\downarrow}-v_k\abar_{\uparrow}. \label{eq:YY}
\end{align}
The $k$-dependent hybridization functions in Eq.(\ref{eq:gX}) and Eq.(\ref{eq:gY}) are defined as
\begin{align}
\Delta_k(\tau, \tau') = \Delta_{-k}(\tau, \tau') = -\im V_k^2 D_{\xi_k}(\tau, \tau'),
\end{align}
which are related to $\Delta(\tau,\tau')$ as
\begin{align}
\Delta(\tau, \tau') = \int_{\xi}\dd\xi \sum_k \Delta_k(\tau, \tau') \delta(\xi-\xi_k),
\end{align}
with the dispersion relation $\xi_k$ is defined in Eq.(\ref{eq:dispersion}).
Now by substituting Eqs.(\ref{eq:XX},\ref{eq:YY},\ref{eq:gX},\ref{eq:gY}) into Eq.(\ref{eq:IFdecomposen}), 
and defining $\boldpsi(\tau) = \left[\begin{array}{cc} a_{\uparrow}(\tau) & \abar_{\downarrow}(\tau) \end{array}\right]$ and $\boldpsibar(\tau) = \left[\begin{array}{cc} \abar_{\uparrow}(\tau) & a_{\downarrow}(\tau) \end{array}\right]$,
we can obtain the final analytic expression of the Feynman-Vernon IF for the SAIM in a compact form:
\begin{align}\label{eq:IFsc}
\gI[\boldabar\bolda] = \exp\left[-\int_{\contour}\dd\tau\int_{\contour}\dd\tau'\boldpsibar(\tau)\boldDelta(\tau-\tau')\boldpsi^T(\tau') \right].
\end{align}
Here $\boldpsi^T$ means the transpose of $\boldpsi$, and $\boldDelta(\tau)= \left[ \begin{array}{cc} \Delta_{\uparrow\uparrow}(\tau) &  \Delta_{\uparrow\downarrow}(\tau) \\
 \Delta_{\downarrow\uparrow}(\tau) &  \Delta_{\downarrow\downarrow}(\tau) 
 \end{array} \right]$ is a $2\times 2$ matrix function defined as
\begin{subequations}\label{eq:hybridization-functions}
\begin{align}
      &\Delta_{\uparrow\uparrow}(\tau)=\int\dd{\varepsilon}[J_{uu}(\varepsilon)D_{\xi}(\tau)-J_{vv}(\varepsilon)D_{\xi}(-\tau)]; \\
      &\Delta_{\uparrow\downarrow}(\tau)=\int\dd{\varepsilon}J_{uv}(\varepsilon)[D_{\xi}(\tau)+D_{\xi}(-\tau)]; \\
      &\Delta_{\downarrow\uparrow}(\tau)=\int\dd{\varepsilon}J_{vu}(\varepsilon)[D_{\xi}(\tau)+D_{\xi}(-\tau)]; \\
      &\Delta_{\downarrow\downarrow}(\tau)=\int\dd{\varepsilon}[J_{vv}(\varepsilon)D_{\xi  }(\tau)-J_{uu}(\varepsilon)D_{\xi}(-\tau)],
\end{align}
\end{subequations}
where we have used the four effective BSFs
\begin{subequations}\label{eq:Js}
\begin{align} 
&J_{uu}(\varepsilon) = J(\varepsilon)u^*(\varepsilon)u(\varepsilon); \\
&J_{uv}(\varepsilon) = J(\varepsilon)u^*(\varepsilon)v(\varepsilon); \\
&J_{vu}(\varepsilon) = J(\varepsilon)v^*(\varepsilon)u(\varepsilon); \\
&J_{vv}(\varepsilon) = J(\varepsilon)v^*(\varepsilon)v(\varepsilon).
\end{align}
\end{subequations}
From Eq.(\ref{eq:IFsc}), we can see that the Grassmann trajectories of the two spin species are indeed coupled, in contrast with Eq.(\ref{eq:IFdecompose}).
The four functions in $\boldDelta(\tau)$ are not independent. In fact, from the definition in Eqs.(\ref{eq:hybridization-functions}), we can see that $\Delta_{\uparrow\uparrow}(\tau) = -\Delta_{\downarrow\downarrow}(-\tau)$ in general, and $\Delta_{\uparrow\downarrow}(\tau) = \Delta_{\downarrow\uparrow}(\tau)$ if $\scgap_k$ is real.
For $\scgap_k = 0$, we have $u_k=1$, $v_k=0$ from Eq.(\ref{eq:uv}), then $J_{uu}(\varepsilon) =J(\varepsilon)$ and $J_{vv}(\varepsilon)=J_{uv}(\varepsilon) =J_{vu}(\varepsilon) = 0$, as a result we have $\Delta_{\uparrow\downarrow}(\tau)=\Delta_{\downarrow\uparrow}(\tau)=0$ and Eq.(\ref{eq:IFsc}) naturally reduces to Eq.(\ref{eq:IFdecompose}) and Eq.(\ref{eq:IFnormal}).

\subsection{The QuAPI scheme for superconducting baths}
The form of Eq.(\ref{eq:hybridization-functions}) inspires a natural way to reuse the QuAPI scheme for the normal bath to discretize the $\boldDelta(\tau)$ for superconducting baths, which is shown in the following.
Assuming that for the case of normal bath, one has implemented the QuAPI scheme which accepts a BSF $J(\varepsilon)$ plus a nontrivial dispersion relation $\xi$, and outputs the discretized hybridization function in Eq.(\ref{eq:quapi}).
Then by taking the four effective BSFs $J_{uu}(\varepsilon)$, $J_{uv}(\varepsilon)$, $J_{vu}(\varepsilon)$, $J_{vv}(\varepsilon)$ in Eqs.(\ref{eq:Js}) plus the dispersion relation in Eq.(\ref{eq:dispersion}) as the inputs, one can obtain the four discretized hybridization functions $\Delta_{uu,i_{\zeta}j_{\zeta'}} $, $\Delta_{uv,i_{\zeta}j_{\zeta'}} $, $\Delta_{vu,i_{\zeta}j_{\zeta'}} $, $\Delta_{vv,i_{\zeta}j_{\zeta'}} $ by using the normal QuAPI scheme for four times. With these four hybridization functions, one can straightforwardly obtain the discretized hybridization functions corresponding to Eqs.(\ref{eq:hybridization-functions}) as
\begin{subequations}\label{eq:dis-hybridization-functions}
\begin{align}
&\Delta_{\uparrow\uparrow,i_{\zeta}j_{\zeta'}} = \Delta_{uu,i_{\zeta}j_{\zeta'}} - \Delta_{vv,i_{\zeta}j_{\zeta'}}^T; \\
&\Delta_{\uparrow\downarrow,i_{\zeta}j_{\zeta'}} = \Delta_{uv,i_{\zeta}j_{\zeta'}} + \Delta_{uv,i_{\zeta}j_{\zeta'}}^T; \\
&\Delta_{\downarrow\uparrow,i_{\zeta}j_{\zeta'}} = \Delta_{vu,i_{\zeta}j_{\zeta'}} + \Delta_{vu,i_{\zeta}j_{\zeta'}}^T; \\
&\Delta_{\downarrow\downarrow,i_{\zeta}j_{\zeta'}} = \Delta_{vv,i_{\zeta}j_{\zeta'}} - \Delta_{uu,i_{\zeta}j_{\zeta'}}^T,
\end{align}
\end{subequations}
which will be used as the input of our Nambu-GTEMPO method.

\subsection{The partial IF algorithm for superconducting baths}
\begin{figure}
  \includegraphics[width=0.95\columnwidth]{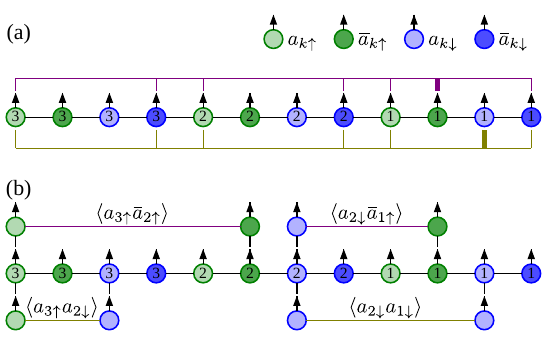} 
  \caption{(a) A schematic illustration of the partial IF algorithm to build the Feynman-Vernon IF as a GMPS for the SAIM, where the imaginary contour has been used as example. The green and blue solid circles represent the GVs for the upper and lower spins respectively, the lighter and darker circles with the same color represent a conjugate pair of GVs. Each horizontal line with legs acting on the circles represent a partial IF, where the thicker leg corresponds to the GV shared by all the terms in the partial IF. Thus, the upper and lower horizontal lines mean two partial IFs that correspond to the first and second terms in the product of Eq.(\ref{eq:partialIFsc}), respectively. (b) Calculating the four four types of imaginary-time correlation functions as defined in Eqs.(\ref{eq:gfimag}), based on the ADT.
    }
    \label{fig:demo}
\end{figure}

After discretization using the QuAPI scheme, one could obtain the discrete IF 
\begin{align}\label{eq:disIFsc}
\gI[\boldabar\bolda] = \exp\left(-\sum_{i_{\zeta}, j_{\zeta'}} \boldpsibar_{i_{\zeta}} \boldDelta_{i_{\zeta} j_{\zeta'}}\boldpsi^T_{j_{\zeta'}} \right).
\end{align}
Each term in the exponent can be explicitly written as the summation of four terms as
\begin{align}
\boldpsibar_{i_{\zeta}} \boldDelta_{i_{\zeta} j_{\zeta'}}\boldpsi^T_{j_{\zeta'}} =& \abar_{\uparrow,i_{\zeta}}\Delta_{\uparrow\uparrow, i_{\zeta}j_{\zeta'}}a_{\uparrow, j_{\zeta'}} + \abar_{\uparrow,i_{\zeta}}\Delta_{\uparrow\downarrow, i_{\zeta}j_{\zeta'}}\abar_{\downarrow, j_{\zeta'}} + \nonumber \\
& a_{\downarrow,i_{\zeta}}\Delta_{\downarrow\uparrow, i_{\zeta}j_{\zeta'}}a_{\uparrow, j_{\zeta'}} + a_{\downarrow,i_{\zeta}}\Delta_{\downarrow\downarrow, i_{\zeta}j_{\zeta'}}\abar_{\downarrow, j_{\zeta'}}.
\end{align}
Now it is straightforward to generalize the partial IF algorithm to this case: by regrouping the terms in Eq.(\ref{eq:disIFsc}) that share a common GV, one obtains
\begin{align}\label{eq:partialIFsc}
\gI[\boldabar\bolda] =& \prod_{i_{\zeta}} \exp\left[\sum_{j_{\zeta'}} \abar_{\uparrow, i_{\zeta}}(\Delta_{\uparrow\uparrow, i_{\zeta}j_{\zeta'}}a_{\uparrow, j_{\zeta'}} + \Delta_{\uparrow\downarrow, i_{\zeta}j_{\zeta'}}\abar_{\downarrow, j_{\zeta'}}) \right] \times \nonumber \\
& \exp\left[\sum_{j_{\zeta'}}a_{\downarrow,i_{\zeta}}(\Delta_{\downarrow\uparrow, i_{\zeta}j_{\zeta'}}a_{\uparrow, j_{\zeta'}} + \Delta_{\downarrow\downarrow, i_{\zeta}j_{\zeta'}}\abar_{\downarrow, j_{\zeta'}}) \right].
\end{align}
We can see that for each $i_{\zeta}$, there are two partial IFs: one shares the common GV $\abar_{\uparrow, i_{\zeta}}$ (i.e., the first line of Eq.(\ref{eq:partialIFsc})), and the other shares the common GV $a_{\downarrow, i_{\zeta}}$ (i.e., the second line of Eq.(\ref{eq:partialIFsc})). As a result, $\gI[\boldabar\bolda]$ can be built as a GMPS by multiplying $2M$ such partial IFs together. 
After each GMPS multiplication, we need to compress the resulting GMPS to keep a fixed bond dimension. In our implementation we set a fixed bond dimension $\chi$ for the bond truncation, namely we keep only the $\chi$ largest singular values for each bond of the GMPS after compression.
The partial IF algorithm for superconducting baths is schematically illustrated in Fig.~\ref{fig:demo}(a), where we have used the imaginary contour as an example. Formally, the central difference between the partial IF algorithm for superconducting baths and that for normal baths is that the GVs corresponding to the two spins are directly coupled in the former case. 
As a result, the IF for both spin species has to be built as a single GMPS in the former case, while they can be built as two separate GMPSs in the latter case. However, in terms of computational costs, the former case is not necessarily harder than the latter, as the coupling between two species may effectively decrease the bond dimension (if there is no coupling between the two species, 
then the bond dimension of the IF for both species is generally the product the bond dimensions for each species~\cite{ChenGuo2024b}).

\subsection{Calculating observables for the SAIM}
Now we show the explicit expressions of the observables we will calculate in the numerical examples of this work, based on the obtained $\gA[\boldabar \bolda]$. 
In our numerical simulations, we will focus on real $\scgap_k$, although our method directly applies for complex $\scgap_k$, we will also fix $\rhoimp$ to be the vacuum state, i.e., $\rhoimp=\vert0 \rangle\langle 0\vert$.
For imaginary-time evolution, we will calculate the following correlation functions: 
\begin{subequations}\label{eq:gfimag}
\begin{align}
-\Gop_{\uparrow\uparrow}(j\delta\tau) &= \langle \aop_{\uparrow}(j\delta\tau)\adop_{\uparrow} \rangle_{\rm eq} \nonumber \\ 
&= \Zimp^{-1}(\beta)\int\mea[\boldabar \bolda] a_{\uparrow, j_o}\abar_{\uparrow, 0_o} \gA[\boldabar \bolda]; \label{eq:gfimag1} \\
\Gop_{\uparrow\downarrow}(j\delta\tau)  &= \langle \aop_{\uparrow}(j\delta\tau)\aop_{\downarrow} \rangle_{\rm eq} \nonumber \\ 
&= \Zimp^{-1}(\beta)\int\mea[\boldabar \bolda] a_{\uparrow, j_o}a_{\downarrow, 0_o} \gA[\boldabar \bolda]; \label{eq:gfimag2} \\
-\Gop_{\downarrow\uparrow}(j\delta\tau)  &= \langle \aop_{\downarrow}(j\delta\tau)\aop_{\uparrow} \rangle_{\rm eq} \nonumber \\ 
&=\Zimp^{-1}(\beta)\int\mea[\boldabar \bolda] a_{\downarrow, j_o}a_{\uparrow, 0_o} \gA[\boldabar \bolda]; \\
-\Gop_{\downarrow\downarrow}(j\delta\tau) &= \langle \aop_{\downarrow}(j\delta\tau)\adop_{\downarrow} \rangle_{\rm eq} \nonumber \\ 
&= \Zimp^{-1}(\beta)\int\mea[\boldabar \bolda] a_{\downarrow, j_o}\abar_{\downarrow, 0_o} \gA[\boldabar \bolda].
\end{align}
\end{subequations}
Here $\delta\tau$ is the imaginary-time step size, $\langle\cdots\rangle_{\rm eq}$ means the expectation value over the thermal state $\exp(-\beta\Hop)$ of the impurity and bath, and we have used $\Zimp(\beta)$ to stress that it is the impurity partition function calculated on the imaginary contour. These correlation functions are the central quantities to calculate in Nambu-DMFT iterations on the imaginary contour (see Appendix.~\ref{app:dmft}).
For real $\scgap$,  they satisfy the symmetry relations $\Gop_{\uparrow\uparrow}(\tau)=\Gop_{\downarrow\downarrow}(\tau)$ and $\Gop_{\uparrow\downarrow}(\tau) = \Gop_{\downarrow\uparrow}(\tau) $ due to the spin symmetry.

For non-equilibrium real-time evolution, we will calculate the following correlation functions:
\begin{subequations}\label{eq:gfreal}
\begin{align}
G_{\uparrow\bar{\uparrow}\bullet}(j\delta t) &=\Zimp^{-1}(t)\Tr\left[\aop_{\uparrow}(j\delta t)\adop_{\uparrow}\rhoop_0\right] \nonumber \\
&= \Zimp^{-1}(t)\int\mea[\boldabar \bolda] a_{\uparrow, j_+}\abar_{\uparrow, 0_+} \gA[\boldabar \bolda]; \label{eq:gfreal1} \\
G_{\bar{\uparrow}\uparrow\bullet}(j\delta t)  &=\Zimp^{-1}(t)\Tr\left[\adop_{\uparrow}(j\delta t)\aop_{\uparrow}\rhoop_0\right] \nonumber \\
&= \Zimp^{-1}(t)\int\mea[\boldabar \bolda] \abar_{\uparrow, 0_+}a_{\uparrow, j_+} \gA[\boldabar \bolda]; \label{eq:gfreal2}\\
G_{\uparrow\downarrow\bullet}(j\delta t)  &=\Zimp^{-1}(t)\Tr\left[\aop_{\uparrow}(j\delta t)\aop_{\downarrow}\rhoop_0\right] \nonumber \\
&= \Zimp^{-1}(t)\int\mea[\boldabar \bolda] a_{\uparrow, j_+}a_{\downarrow, 0_+} \gA[\boldabar \bolda]; \label{eq:gfreal3}\\
G_{\uparrow\bullet\downarrow}(j\delta t) &=\Zimp^{-1}(t)\Tr\left[\aop_{\uparrow}(j\delta t)\rhoop_0 \aop_{\downarrow}\right] \nonumber \\
&= \Zimp^{-1}(t)\int\mea[\boldabar \bolda] a_{\uparrow, 0_+}a_{\downarrow, j_-} \gA[\boldabar \bolda]. \label{eq:gfreal4}
\end{align}
\end{subequations}
Here $\delta t$ means the real-time step size and we have used $\Zimp(t)$ with $t=j\delta t$ to stress that it is the impurity partition function calculated on the Keldysh contour. 
We note that the arrows in Eq.(\ref{eq:gfreal}) stand for the spins of the annihilation and creation (if there is a bar above the arrow) operators, while the bullet stands for $\rhoop_0$. Therefore Eqs.(\ref{eq:gfreal1}, \ref{eq:gfreal2}, \ref{eq:gfreal3}) have the same form as the usual greater and lesser Green's functions.
However, the correlation function defined in Eq.(\ref{eq:gfreal4}) has a very special order (noticing the corresponding time step index $j_-$ in the second line), which is different from the usual Green's functions and is particularly used to demonstrate the flexibility of our method to calculate any impurity observables.
Since we only consider $\rhoimp$ to be the vacuum state, we further have $G_{\bar{\uparrow}\uparrow\bullet}(j\delta t) = G_{\uparrow\downarrow\bullet}(j\delta t) = 0$ ($G_{\uparrow\bullet\downarrow}(j\delta t)$ does not vanish as $\aop_{\downarrow}$ acts on the right of $\rhoop_0$).
In addition, for real-time evolution we will also calculate the probability of the impurity on the four impurity states, i.e., $\vert 0\rangle$ (empty), $\vert \uparrow\rangle$ (spin up), $\vert \downarrow\rangle$ (spin down), $\vert \uparrow\downarrow\rangle$ (double occupation):
\begin{subequations}\label{eq:populations}
\begin{align}
&P_{ 0}(j\delta t) = \Zimp^{-1}(t)\int\mea[\boldabar \bolda] a_{\uparrow, j_+} \abar_{\uparrow, j_+} a_{\downarrow, j_+} \abar_{\downarrow, j_+} \gA[\boldabar \bolda];\\
&P_{ \uparrow}(j\delta t) = \Zimp^{-1}(t)\int\mea[\boldabar \bolda] a_{\uparrow, j_+} \abar_{\uparrow, j_-} a_{\downarrow, j_+} \abar_{\downarrow, j_+} \gA[\boldabar \bolda];\\
&P_{ \downarrow}(j\delta t) = \Zimp^{-1}(t)\int\mea[\boldabar \bolda] a_{\uparrow, j_+} \abar_{\uparrow, j_+} a_{\downarrow, j_+} \abar_{\downarrow, j_-} \gA[\boldabar \bolda];\\
&P_{ \uparrow\downarrow}(j\delta t) = \Zimp^{-1}(t)\int\mea[\boldabar \bolda] a_{\uparrow, j_+} \abar_{\uparrow, j_-} a_{\downarrow, j_+} \abar_{\downarrow, j_-} \gA[\boldabar \bolda],
\end{align}
\end{subequations}
which satisfies $P_{ \uparrow}(j\delta t) = P_{ \downarrow}(j\delta t)$ due to the spin symmetry.

\section{Numerical results}\label{sec:results}
In this section we will present our numerical results for the superconducting Anderson impurity model. 
We will first perform converged Nambu-DMFT iterations on the imaginary contour, and benchmark our results against CTQMC results, and then we will perform non-equilibrium real-time evolution and study the time evolution of the probability on the impurity states. 
The validation of the Nambu-DMFT method against exactly solvable cases is shown in Appendix.~\ref{app:validate}.



\subsection{Imaginary-time DMFT}\label{sec:dmft}

\begin{figure}
  \includegraphics[]{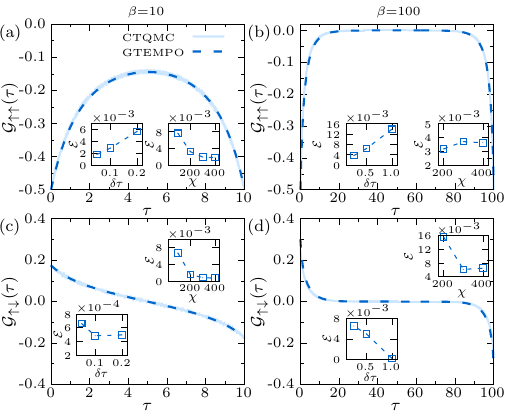} 
  \caption{(a,c) The imaginary-time correlation functions $\Gop_{\uparrow\uparrow}(\tau)$ (a) and $\Gop_{\uparrow\downarrow}(\tau)$ (c) as functions of the imaginary time $\tau$ for the SAIM with $\beta=10$, where the blue dashed lines are the Nambu-GTEMPO results calculated with $\chi=400$ and $\delta\tau=0.05$, and the solid lines with the same color are the corresponding CTQMC results. The two insets in each panel show the mean errors between Nambu-GTEMPO and CTQMC as functions of $\delta\tau$ with fixed $\chi=400$ (left inset), and as functions of $\chi$ with fixed $\delta\tau=0.05$ (right inset), respectively. (b,d) similar to (a,c), but for the SAIM with $\beta=100$. The Nambu-GTEMPO results in the main panels are calculated with $\chi=400$ and $\delta\tau=0.25$. The two insets show the mean errors between Nambu-GTEMPO and CTQMC as functions of $\delta\tau$ with fixed $\chi=400$ (left inset), and as functions of $\chi$ with fixed $\delta\tau=0.25$ (right inset), respectively. We have used $U=-1$ and $\varepsilon_d=-U/2$ in these simulations. 
    }
    \label{fig:siamimag}
\end{figure}

Here we showcase the application of Nambu-GTEMPO to the superconducting states of the one-band attractive Hubbard model within DMFT. 
We will choose $U=-1$ and set the system half-filled ($\varepsilon_d=-U/2$).
We focus on the attractive-$U$ case because the single-band attractive Hubbard model naturally supports an on-site, intra-orbital, spin-singlet $s$-wave superconducting state, which can be described within single-site DMFT in the Nambu formalism. Although the half-filled negative-$U$ Hubbard model can be mapped to a positive-$U$ model in the absence of superconductivity, the attractive-$U$ model provides the simplest self-consistent test bed for benchmarking a Nambu impurity solver. By contrast, superconductivity in the repulsive-$U$ Hubbard model is expected to have predominantly $d$-wave character and therefore requires cluster DMFT, typically on at least a $2\times 2$ plaquette (for example, see Ref.~\cite{Sordi2012}), which is beyond the scope of the present Nambu-GTEMPO calculations.

\begin{figure}
  \includegraphics[]{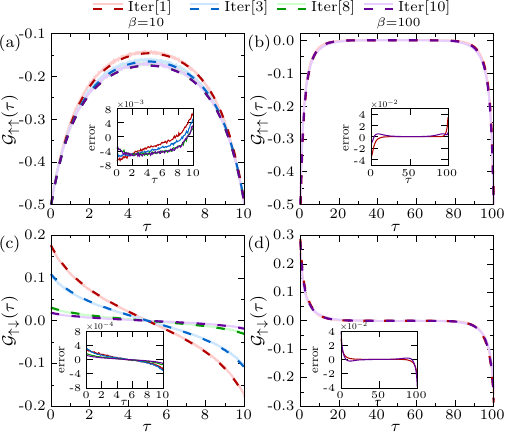} 
  \caption{(a,c) The imaginary-time correlation functions $\Gop_{\uparrow\uparrow}(\tau)$ (a) and $\Gop_{\uparrow\downarrow}(\tau)$ (c) as functions of the imaginary time $\tau$ for the SAIM with $\beta=10$ during $10$ DMFT iterations, where the dashed lines in different colors are the Nambu-GTEMPO results calculated with $\chi=200$ and $\delta\tau=0.1$ at the selected iterations, and the solid lines with the same colors are the corresponding CTQMC results. (b,d) similar to (a, c), but for the SAIM with $\beta=100$, where the Nambu-GTEMPO results are calculated with $\chi=300$ and $\delta\tau=0.5$ instead. The insets show the errors between Nambu-GTEMPO and CTQMC at the selected iterations, in correspondence with the main panels.
  We have used $U=-1$ and $\varepsilon_d=-U/2$ in these simulations.
    }
    \label{fig:siamimagdmft}
\end{figure}

In Fig.~\ref{fig:siamimag} and Fig.~\ref{fig:siamimagdmft}, we study the imaginary-time evolution of the SAIM and compare our results to CTQMC, where in Fig.~\ref{fig:siamimag} we focus on the error analysis against the two hyperparameters $\chi$ and $\delta\tau$ in a single DMFT iteration, and in Fig.~\ref{fig:siamimagdmft} we perform converged DMFT iterations with fixed $\chi$ and $\delta\tau$. Here due to the particle-hole symmetry of the SAIM, the CTQMC is free from the sign problem and thus provides a reliable benchmark. The details of the DMFT iterations using Nambu-GTEMPO as the impurity solver, can be found in Appendix.~\ref{app:dmft}. The starting point for the DMFT, using both the CTQMC and Nambu-GTEMPO solvers, is an anomalous hybridization function with a small anomalous seed, which is also specified in Appendix.~\ref{app:dmft}.

In Fig.~\ref{fig:siamimag}(a,c), we plot $\Gop_{\uparrow\uparrow}(\tau)$ and $\Gop_{\uparrow\downarrow}(\tau)$ as functions of the imaginary time $\tau$ for $\beta=10$ (we note that the room temperature is around $\beta=40$ in our units), where the blue dashed lines are the Nambu-GTEMPO results and the solid lines with the same color are the corresponding CTQMC results. In Fig.~\ref{fig:siamimag}(b,d), we plot the same quantities as in Fig.~\ref{fig:siamimag}(a,c), but for $\beta=100$ instead. For both values of $\beta$, we can see a good match between the Nambu-GTEMPO results and the CTQMC results. In the insets, we show the mean errors as functions of $\delta\tau$ (left inset) and of $\chi$ (right inset) respectively. From the insets of Fig.~\ref{fig:siamimag}(a,c), we can see that the mean errors generally decrease for smaller $\delta\tau$ or larger $\chi$, which is as expected. However, the right inset of Fig.~\ref{fig:siamimag}(b) shows a slight increase of $\mathcal{E}$ as $\chi$ increases from $200$ to $300$, which is likely a coincidence due to the interplay between the two error sources. In addition, the bottom left inset of Fig.~\ref{fig:siamimag}(d) shows a increase of $\mathcal{E}$ as $\delta\tau$ decreases. This is likely because that the errors of $\Gop_{\uparrow\downarrow}(\tau)$ are mostly located at the boundaries (which has also been observed in Ref.~\cite{ChenGuo2024b} for the standard AIM), and if we use a large $\delta\tau$, very few data points are selected from the boundaries, which makes the mean error smaller (the mean error for $\delta\tau=1$ is already of the order $10^{-4}$, more than two orders of magnitude smaller than the time step size we have used, illustrating the higher accuracy of the middle data points in the Nambu-GTEMPO results).

In Fig.~\ref{fig:siamimagdmft}, we further perform converged imaginary-time DMFT iterations on the Bethe lattice with hopping $\nu=D/2$ for both $\beta=10$ (the left panels) and $\beta=100$ (the right panels). In Fig.~\ref{fig:siamimagdmft}(a,c), we plot $\Gop_{\uparrow\uparrow}(\tau)$ and $\Gop_{\uparrow\downarrow}(\tau)$ at different DMFT iterations, where the dashed lines are the Nambu-GTEMPO results calculated with $\chi=200$ and $\delta\tau=0.1$ and the solid lines are the corresponding CTQMC results. In Fig.~\ref{fig:siamimagdmft}(b,d) we plot the same quantities as in Fig.~\ref{fig:siamimagdmft}(a,c), but for $\beta=100$, and we have used $\chi=300$ and $\delta\tau=0.5$ in the Nambu-GTEMPO calculations (we have only shown the first and last DMFT iterations in this case as the DMFT calculation quickly converges). 
We can see a step by step matching between Nambu-GTEMPO and CTQMC.
In the insets, we show the difference (error) between the Nambu-GTEMPO and CTQMC results at each iteration, from which we can see that the errors generally become smaller at larger iterations, until convergence has been reached.
These results demonstrate the potential of our Nambu-GTEMPO method as an powerful candidate quantum impurity solver to study the superconducting states within DMFT.
Compared to CTQMC, the advantages of Nambu-GTEMPO on the imaginary contour include: (1) the results are free of sampling noises; (2) it is free of the sign problem, essentially because that the Nambu-GTEMPO method is deterministic, similar to usual MPS based algorithms. 
However, currently the efficiency of Nambu-GTEMPO is not comparable to CTQMC. Taking the calculations in Eq.(\ref{fig:siamimagdmft}) for $\beta=100$ as an example, it takes about $10$ hours per DMFT iteration using Nambu-GTEMPO on a single thread of a CPU with $2.4$ GHz frequency (Multi-process parallelization of GTEMPO is challenging, as a general feature of tensor network algorithms, but it could be promising to use the high-performance, high-bandwidth architectures, such as GPU, to accelerate GTEMPO), while for CTQMC it takes only $15$ minutes to obtain a total of $128\times10^6$ samples per DMFT iteration using $128$ CPU cores, each with $3.1$ GHz frequency.
The ideal application scenario for our method is where CTQMC suffers from a severe sign problem, such as off-diagonal impurity-bath coupling, or real-time evolution. The latter scenario where be studied in the following. 

\subsection{Real-time evolution}

\begin{figure}
  \includegraphics[width=\columnwidth]{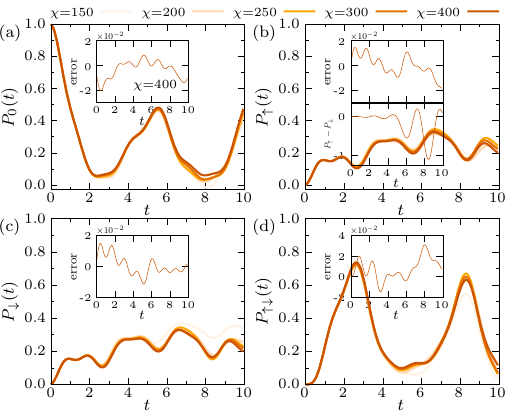} 
  \caption{Time evolution of the probability of the four impurity states $\vert 0\rangle$ (a), $\vert \uparrow\rangle$ (b), $\vert\downarrow\rangle$ (c) and $\vert\uparrow\downarrow\rangle$ (d), for the SAIM. The orange solid lines from lighter to darker are the Nambu-GTEMPO results calculated with different $\chi$s from small to large, with fixed $\delta t=0.05$. In the insets, we show the errors between the Nambu-GTEMPO results calculated using $\delta t=0.05$ and $\delta t=0.1$, with fixed $\chi=400$. In (b), the lower inset shows the time evolution of $P_{\uparrow}(t)-P_{\downarrow}(t)$, calculated with $\delta t=0.05$ and $\chi=400$. We have used $U=-1$, $\varepsilon_d=-U/2$ and $\beta=10$ in these simulations.
    }
    \label{fig:siamneq}
\end{figure}

Now we study the real-time evolution of the SAIM, which is a very challenging task for non-equilibrium CTQMC due to the dynamical sign problem. We consider the attractive-U case and choose the same parameters for the impurity as in Sec.~\ref{sec:dmft}.
We consider the commonly used semi-circular BSF:
\begin{align}\label{eq:semicircular}
J(\varepsilon) = \frac{2}{\pi D}\sqrt{1-\frac{\varepsilon^2}{D^2}},
\end{align}
where we set $D=1$ and use it as the unit.
In addition, we choose $\scgap_k = \scgap =1$. This relatively large bath pairing amplitude is chosen for benchmarking purposes, since it makes the superconducting pairing effects clearly visible.

In this simulation, we consider a quench scenario where the impurity with an attractive interaction suddenly couples to a BCS
bath with the superconducting gap size $2\scgap=2$. We assume that the impurity is initially in the empty state $\vert 0\rangle$ and the bath is initially in the vacuum state.
In Fig.~\ref{fig:siamneq}, we show the time evolution of probability of the four impurity states defined in Eqs.(\ref{eq:populations}), as functions of the real time $t$ in the four panels respectively. 
The orange solid lines from lighter to darker are the Nambu-GTEMPO results calculated with different $\chi$s ranging from $150$ to $400$, with fixed $\delta t=0.05$. We can see that for $\chi\geq 200$, the different lines already become very close to each other. 
We note that $P_{\uparrow}(t)$ should be exactly equal to $P_{\downarrow}(t)$ due to spin symmetry. However, from Fig.(\ref{fig:siamneq})(b,c) we can see that this symmetry is not exactly preserved, especially for $\chi=150$. The reason is that we have not explicitly encoded this symmetry 
into the GMPS (which could be explored in future investigations), and that for small $\chi$ the Nambu-GTEMPO calculations are not accurate enough.
In the insets, we plot the errors between the Nambu-GTEMPO results calculated using $\delta t=0.05$ and $\delta t=0.1$, with fixed $\chi=400$. We can see that the errors are generally less than $2\times 10^{-2}$. 
Interestingly, for the real-time evolution of a similar impurity problem but coupled to a normal bath, it has been shown that the bond dimension of the IF for each spin species is set to $80$ to obtain accurate results~\cite{SunGuo2025c} (which effectively amounts to a bond dimension $6400$ for the IF of both species), 
while here we have only used $\chi=400$ for the IF for both species to obtain fairly well converged results, therefore the (large) coupling between the two spin species within the superconducting bath can indeed decrease the overall bond dimension of the IF for the two species.
These results show the potential of our method to be used as an impurity solver in non-equilibrium DMFT.

From a physical picture perspective, we can identify two charge-transfer channels from the real-time dynamics of the impurity.
In Fig.~\ref{fig:siamneq}(a,d), we observe a Cooper pair tunneling dynamics dominated by the period $T_p \sim 5.8$.
The corresponding energy $\omega_p = 2\pi/T_p \sim 1.1$ lies below the superconducting
gap $2\scgap=2$. This in-gap energy scale is a hallmark of Andreev tunneling,
where a Cooper pair tunnels as an intact unit between the superconducting bath and the
normal-state impurity, $|0\rangle \leftrightarrow |\!\uparrow\downarrow\rangle$, without
dissociating into Bogoliubov quasiparticles. If we set the initial state of the impurity in the
particle--hole symmetric state, given our half-filling Hamiltonian, the two states
$|0\rangle$ and $|\!\uparrow\downarrow\rangle$ would be degenerate. Even though we now start
from an empty state, the Andreev-mediated hybridization gives rise to a coherent oscillation
with a frequency inherited from the bath.
On the contrary, in Fig.~\ref{fig:siamneq}(b,c) we observe a faster dynamics of the single-particle hopping
with period $T_s \sim 2.4$. The corresponding energy $\omega_s=2\pi/T_s\sim 2.5$ exceeds
$2\scgap$, indicating that the process involves real Bogoliubov quasiparticle
excitations, in which a Cooper pair in the bath dissociates. One electron
enters the impurity as a single-occupancy state $|\!\uparrow\rangle$ or $|\!\downarrow\rangle$,
while the other escapes as a Bogoliubov quasiparticle at an energy cost $\omega_s>2\scgap$.

We note that in this simulation of the quench dynamics we have not yet reached equilibrium, which likely requires a much longer time evolution. 
The main reason why we only simulate the dynamics till $t=10$ is that both the space and time costs of our method scales linearly 
against $t$ under a fixed bond dimension (the cost of MPS methods generally scale linearly with the total number of sites), and we will face an MPS with 
thousands of sites for $t\sim 100$ which would be very demanding. 
This issue, however, could in principle be largely alleviated by using high-performance computing devices such as GPU (we only used a single thread of CPU in our current implementation). 
Due to this limitation (finite time window and frozen bath), 
we are not yet
capable of studying the coherent Higgs mode induced by the hybridization quench between the
impurity and the superconducting bath.  A fully
self-consistent real-time DMFT calculation, where $\scgap$ evolves dynamically
in response to the impurity dynamics, would provide access to the
collective Higgs responses and their interplay with the Andreev and Cooper pair-breaking
dynamics identified here.

In addition, we note that there is a sharp difference in the error behavior during the time evolution between our method and the conventional tensor network based time evolution methods. In conventional methods such as the time-evolving MPS method~\cite{Schollwock2011,PaeckelHubig2019}, the error is generally $0$ at the beginning and accumulates during the time evolution, while in our method the error is distributed over the entire time axis, with no particular bias towards short or long times (there is a finite error even at $t=0$ as can be seen from Fig.~\ref{fig:siamneq}). This is because our method is formulated from the IF obtained by integrating out the bath over the entire time interval, which is then discretized uniformly using the QuAPI scheme. Moreover,  we have not explicitly preserved the causal relation during our construction of the MPS representation of the IF, as opposed to the original TEMPO, so the errors in the history and future are not disentangled in our approach.

\section{Summary}\label{sec:summary}

In summary, we have proposed an extension of the GTEMPO method for the superconducting Anderson impurity model in which the impurity is coupled to a superconducting fermionic bath with s-wave pairing. A key step in our extension is to use the Bogoliubov transformation to convert the superconducting bath into the form of a normal bath, such that the analytic expression of the Feynman-Vernon influence functional can be derived. 
We validate our method by benchmarking it against the CTQMC results in the non-integrable SAIM by performing converged DMFT iterations on the imaginary-time axis. 
As an illustration of the powerfulness of our method, we apply it to study the non-equilibrium real-time evolution of the SAIM, which is a very challenging task for CTQMC due to the dynamical sign problem. We note that our method can be straightforwardly extended to treat other forms of pairing in both multi-orbital and multi-site QIP. These results demonstrate the flexibility and potential of our Nambu-GTEMPO method as the impurity solver in DMFT and non-equilibrium DMFT to study the superconducting states in both equilibrium and non-equilibrium systems.

An interesting future subject is to apply the Nambu-GTEMPO method to study larger impurity problems. However, a straightforward application of the current approach for a multiple-orbital impurity would result in an exponential growth of the computational cost against the number of orbitals, as demonstrated in vanilla GTEMPO~\cite{ChenGuo2024b}. Nevertheless, it has been illustrated that the cost could be greatly reduced by focusing on only one flavor (or one orbital) and integrating out all the other flavors in the IF before hand~\cite{SunGuo2025b}. This technique can be straightforwardly adopted in Nambu-GTEMPO method.


The implementation of the Nambu-GTEMPO method proposed in this work on the imaginary and Keldysh, as well as the L-shaped Kadanoff contours (calculations on the last contour have not been considered in the numerical examples of this work), could be found at~\cite{GTEMPO} (which is written in the julia language).

\begin{acknowledgments}
We greatly thank Jiyu Chen for helping us to interpret the numerical results for the real-time evolution of the superconducting Anderson impurity model with attractive interaction. This work is supported by National Natural Science Foundation of China under Grant No. 12104328, No.12574403, No. 12174447, No. 22542010, No. 62471478, No. 12305049 and 12474231. Changming Yue also acknowledges supports from the Guangdong Major Project of Basic Research (2025B0303000004),  Guangdong Provincial Quantum Science Strategic Initiative (GDZX2401004, GDZX2501001).   
\end{acknowledgments}

\textit{Data availability statement.} The data that support the findings of this article are openly available \cite{NAMBU-GTEMPO-DATA}.

\appendix

\section{DMFT iterations on the imaginary contour using Nambu-GTEMPO}\label{app:dmft}
As we focus on real $\scgap_k$, it can be seen from
Eq.(\ref{eq:hybridization-functions}) that
$\Delta_{\uparrow\uparrow}(\tau)=-\Delta_{\downarrow\downarrow}(-\tau)$ and $\Delta_{\uparrow\downarrow}(\tau)=\Delta_{\downarrow\uparrow}(\tau)$. Denoting $\Delta_{\uparrow\uparrow}$ as $\Delta^{\mathrm{nor}}$ (the normal hybridization function), and $\Delta_{\uparrow\downarrow}$ as $\Delta^{\mathrm{ano}}$ (the anomalous hybridization function), then the $\boldDelta(\im\omega_n)$ in the frequency space (which is the Fourier transformation of $\boldDelta(\tau)$) can be written as
\begin{align}
  \boldDelta(\im\omega_n)=\mqty[\Delta^{\mathrm{nor}}(\im\omega_n)&\Delta^{\mathrm{ano}}(\im\omega_n)\\
  \Delta^{\mathrm{ano}}(\im\omega_n)&-\Delta^{\mathrm{nor}}(-\im\omega_n)],
\end{align}
where $\omega_n$ is the Matsubara frequency.
We consider the DMFT iterations on the Bethe lattice with the lattice hopping constant $\nu=D/2=0.5$, for which the self-consistency condition takes a very simple form~\cite{YueWerner2022}:
\begin{align}\label{eq:bethe}
  \frac{\boldDelta(\im\omega_n)}{\nu^2}=\sgz\boldGop(\im\omega_n)\sgz=
  \mqty[\Gop^{\mathrm{nor}}(\im\omega_n)&-\Gop^{\mathrm{ano}}(\im\omega_n)\\
  -\Gop^{\mathrm{ano}}(\im\omega_n)&-\Gop^{\mathrm{nor}}(-\im\omega_n)],
\end{align}
where $\sgz = \left[\begin{array}{cc} 1 & 0 \\ 0 & -1\end{array}\right]$, and we have used $\Gop^{\mathrm{nor}}(\tau) = \Gop_{\uparrow\uparrow}(\tau)$, $\Gop^{\mathrm{ano}}(\tau) = \Gop_{\uparrow\downarrow}(\tau)$ with $\Gop_{\uparrow\uparrow}(\tau)$ and $\Gop_{\uparrow\downarrow}(\tau)$ defined in Eq.(\ref{eq:gfimag1}) and Eq.(\ref{eq:gfimag2}), and $\Gop_{\uparrow\uparrow}(\im\omega_n)$ and $\Gop_{\uparrow\downarrow}(\im\omega_n)$ are the Fourier transformation of $\Gop_{\uparrow\uparrow}(\tau)$ and $\Gop_{\uparrow\downarrow}(\tau)$ respectively.
Using Nambu-GTEMPO as the imaginary-time impurity solver and given an initial $\boldDelta_0(\im \omega_n)$ (which can either be directly given, or be obtained by substituting a guess BSF into Eq.(\ref{eq:hybridization}) and then performing the Fourier transformation of the resulting $\boldDelta(\tau)$), we could perform the DMFT iteration as follows: (1) calculating the discretized $\boldDelta(\tau)$ using the inverse Fourier transformation of $\boldDelta(\im \omega_n)$; (2) performing Nambu-GTEMPO calculation to obtain $\gA[\boldabar\bolda]$ using $\boldDelta(\tau)$ as the input, and calculating $\boldGop(\tau)$ using Eqs.(\ref{eq:gfimag}) (here we note that Nambu-GTEMPO only requires the $\boldDelta(\tau)$ as the input, instead of the detailed knowledge of $\Hint$, similar to CTQMC); (3) Performing Fourier transformation of $\boldGop(\tau)$ to obtain $\boldGop(\im \omega_n)$, and then obtain $\boldDelta(\im\omega_n)$ using Eq.(\ref{eq:bethe}), which can be used as the starting point of the next iteration.
In the DMFT iterations of this work, we choose the initial $\boldDelta_0(\im \omega_n)$ as:
\begin{align}
&\Delta^{\rm nor}_0(\im \omega_n) = 2\im\omega_n \nu^2\left[1-\sqrt{1+\frac{1}{\omega_n^2 + (U\scgap_0)^2}} \right]; \\
&\Delta^{\rm ano}_0(\im \omega_n) = -2\nu^2 U\scgap_0\left[1- \sqrt{1+\frac{1}{\omega_n^2 + (U\scgap_0)^2}} \right],
\end{align}
in which we set $U\scgap_0 = 0.2$ and $\nu=0.5$.
Here $U\Delta_0^{\rm sc}$ is only used to initialize the anomalous hybridization function in the DMFT loop. We choose the small value $U\Delta_0^{\rm sc}=0.2$ as an initial anomalous seed. During the DMFT iterations, the anomalous component is determined self-consistently and is enhanced by the attractive interaction.

\section{Additional numerical results to validate Nambu-GTEMPO against exactly solvable cases}\label{app:validate}

In this section, we will validate our Nambu-GTEMPO method in two exactly solvable cases: (1) a toy model where the bath contains a single orbital; and (2) the non-interacting case with $U=0$. 
In the first case, the problem can be exactly solved by directly diagonalizing $\Hop$, while in the second case, the problem can be exactly solved by first discretizing the bath into $N$ modes, 
and then diagonalizing a $4(N+1)\times 4(N+1)$ coefficient matrix (the details of the ED algorithm used for non-interacting fermions is shown in Appendix.~\ref{app:noninteractinged}). 
For all these simulations we only consider a single-step impurity solver and we will fix $\scgap_k = \scgap =1$ for the BCS bath.

\subsection{Toy model with a single-orbital bath}

\begin{figure}
  \includegraphics[]{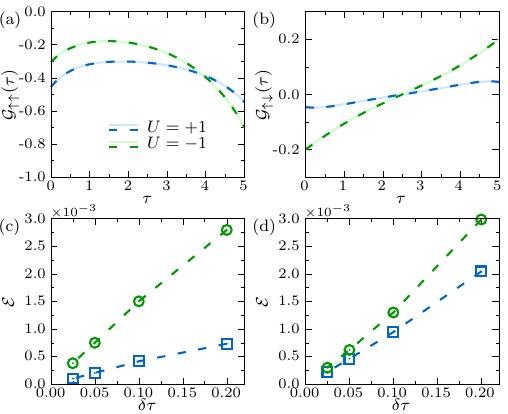} 
  \caption{(a,b) The imaginary-time correlation functions $\Gop_{\uparrow\uparrow}(\tau)$ (a) and $\Gop_{\uparrow\downarrow}(\tau)$ (b) as functions of the imaginary time $\tau$ for the toy model. The blue and green dashed lines are the Nambu-GTEMPO results, calculated with $\chi=30$ and $\delta\tau=0.05$, for $U=1$ and $U=-1$ respectively, while the solid lines with the same colors are the corresponding ED results. (c,d) The mean errors of $\Gop_{\uparrow\uparrow}(\tau)$ (c) and $\Gop_{\uparrow\downarrow}(\tau)$ (d), between Nambu-GTEMPO with $\chi=30$ and ED, as functions of the imaginary-time step size $\delta\tau$, where the blue and green dashed lines are results for $U=1$ and $U=-1$ respectively. We have used $\varepsilon_0 = 1$, $\alpha_0^2=0.5$, $\beta=5$ and $\varepsilon_d=-U/2$ in these simulations. 
    }
    \label{fig:toymodelimag}
\end{figure}

The Hamiltonian we consider for the toy model is
\begin{align}
&\Hbath = \varepsilon_0\sum_{\sigma}\cdop_{\sigma}\cop_{\sigma} + \scgap(\cdop_{\uparrow}\cdop_{\downarrow} + \cop_{\downarrow}\cop_{\uparrow}); \\
&\Hhyb = \alpha_0 \sum_{\sigma}(\bdop_{\sigma}\cop_{\sigma} + \hc),
\end{align}
which corresponds to the BSF
\begin{align}
J(\varepsilon) = \alpha_0^2 \delta(\varepsilon-\varepsilon_0).
\end{align}
In our simulations for the toy model, we will fix $\varepsilon_0 = 1$, $\alpha_0^2=0.5$, and $\varepsilon_d=-U/2$.

We first study the imaginary-time evolution of the toy model for $\beta=5$, and calculate the imaginary-time correlation functions defined in Eq.(\ref{eq:gfimag}). We plot $\Gop_{\uparrow\uparrow}(\tau)$ and $\Gop_{\uparrow\downarrow}(\tau)$ as functions of the imaginary time $\tau$ in Fig.~\ref{fig:toymodelimag}(a) and Fig.~\ref{fig:toymodelimag}(b), respectively ($\Gop_{\downarrow\downarrow}(\tau)$ and $\Gop_{\downarrow\uparrow}(\tau)$ are not shown as they can be obtained from the two correlation functions we have plotted). We have considered both repulsive interaction with $U=1$ (the blue dashed lines) and attractive interaction with $U=-1$ (the green dashed lines), and plot the corresponding ED results (the solid lines with the same color) for comparison. We can see that the Nambu-GTEMPO results generally agree very well with the ED results in all these cases.
In Fig.~\ref{fig:toymodelimag}(c), we plot the mean error $\mathcal{E}$, defined as $\mathcal{E}(\vec{x}, \vec{y}) = \sqrt{||\vec{x}-\vec{y}||^2/L}$ for two vectors $\vec{x}$, $\vec{y}$ of length $L$, between $\Gop_{\uparrow\uparrow}$ calculated by Nambu-GTEMPO and by ED for $U=1$ (the blue dashed line) and $U=-1$ (the green dashed line) respectively. In Fig.~\ref{fig:toymodelimag}(d), we plot the mean error $\mathcal{E}$ between $\Gop_{\uparrow\downarrow}$ calculated by Nambu-GTEMPO and by ED, similar to Fig.~\ref{fig:toymodelimag}(c). We can see that the mean errors decrease almost linearly as $\delta\tau$ decreases. This is expected because that our method is based on a first-order Trotter decomposition (the IF in Eq.(\ref{eq:disIFsc}) is discretized using a first-order QuAPI scheme), and that the time discretization error is essentially the only error source in our method for the toy model: 
the bond dimension of the IF is strictly limited when the bath contains only a single orbital (or more generally a finite number of orbitals)~\cite{SunGuo2025c}. In particular, for the single-orbital bath considered here, we expect that the bond dimension of the MPS representation of the IF is at most $4$ (spin up and down) for imaginary-time evolution, and $16$ for real-time evolution (there are two branches), thus the MPS bond truncation error vanishes as long as the bond dimension $\chi$ used for building the IF is larger than such limit.

\begin{figure}
  \includegraphics[]{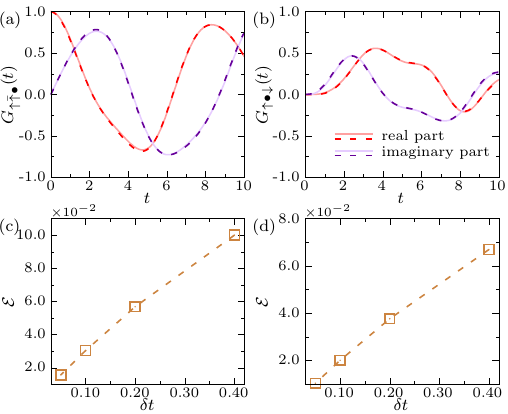} 
  \caption{(a,b) The real-time correlation functions $G_{\uparrow\bar{\uparrow}\bullet}( t)$ (a) and $G_{\uparrow\bullet\downarrow}( t)$ as functions of the real time $t$ for the toy model. The red and purple dashed lines are the real and imaginary parts of the Nambu-GTEMPO results calculated with $\chi=40$ and $\delta t=0.05$, while the solid lines with the same colors are the corresponding ED results. (c,d) The mean error of $G_{\uparrow\bar{\uparrow}\bullet}( t)$ (c) and $G_{\uparrow\bullet\downarrow}( t)$ (d), between Nambu-GTEMPO with $\chi=40$ and ED, as functions of the real-time step size $\delta t$. We have used $\varepsilon_0 = 1$, $\alpha_0^2=0.5$, $\beta=10$, $U=1$ and $\varepsilon_d=-U/2$ in these simulations. 
    }
    \label{fig:toymodelreal}
\end{figure}

In Fig.~\ref{fig:toymodelreal}, we study the real-time evolution of the toy model starting from the uncorrelated impurity-bath initial state in Eq.(\ref{eq:rho0}), for which we set $\beta=10$, $U=1$ and $\varepsilon_d=-U/2$. 
In Fig.~\ref{fig:toymodelreal}(a,b), we plot $G_{\uparrow\bar{\uparrow}\bullet}( t)$ and $G_{\uparrow\bullet\downarrow}( t)$ as functions of the real time $t$ respectively, where we can see that the Nambu-GTEMPO results match very well with ED results. 
In Fig.~\ref{fig:toymodelreal}(c,d), we plot the mean errors of $G_{\uparrow\bar{\uparrow}\bullet}( t)$ and $G_{\uparrow\bullet\downarrow}( t)$ between Nambu-GTEMPO and ED as functions of the real-time step size $\delta t$, where we can see that $\mathcal{E}$ decreases almost linearly as $\delta t$ decreases, similar to the observations in Fig.~\ref{fig:toymodelimag}(c,d).

To this end, we briefly discuss the choices of the two hyperparameters $\chi$ and $\delta t$ in practice. Generally, $\delta t$ should be the smallest one of all the time scales in the problem, as it plays the role of time discretization for solving a differential equation, therefore it should be smaller for larger $U$. For imaginary-time evolution, it has been shown that a larger $\chi$ is required for larger $\beta$~\cite{ChenGuo2024b}, while the relation between $\beta$ and $\chi$ in the real-time evolution is not as clear.


\subsection{The non-interacting case}

\begin{figure}
  \includegraphics[]{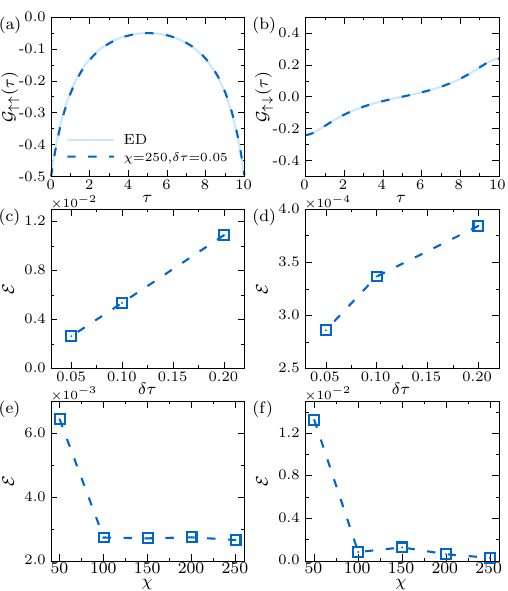} 
  \caption{(a,b) The imaginary-time correlation functions $\Gop_{\uparrow\uparrow}(\tau)$ (a) and $\Gop_{\uparrow\downarrow}(\tau)$ (b) as functions of the imaginary time $\tau$ for the non-interacting case. The blue dashed lines are the Nambu-GTEMPO results calculated with $\chi=250$ and $\delta\tau=0.05$, while the solid lines with the same color are the corresponding ED results. (c,d) The mean errors of $\Gop_{\uparrow\uparrow}(\tau)$ (c) and $\Gop_{\uparrow\downarrow}(\tau)$ (d), between Nambu-GTEMPO with $\chi=250$ and ED, as functions of the imaginary-time step size $\delta\tau$. (e,f) The mean errors of $\Gop_{\uparrow\uparrow}(\tau)$ (e) and $\Gop_{\uparrow\downarrow}(\tau)$ (f), between Nambu-GTEMPO with $\delta \tau=0.05$ and ED, as functions of the bond dimension $\chi$. We have used $\varepsilon_d = 0$, $\beta=10$ in these simulations. }
    \label{fig:toulouseimag}
\end{figure}

\begin{figure}
  \includegraphics[]{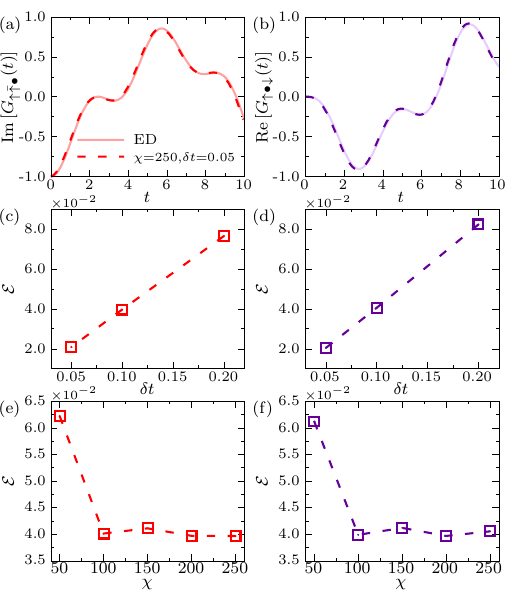} 
  \caption{(a,b) The real-time correlation functions $G_{\uparrow\bar{\uparrow}\bullet}( t)$ (the real part vanishes and is not shown) (a) and $G_{\uparrow\bullet\downarrow}( t)$ (the imaginary part vanishes and is not shown) as functions of the real time $t$ for the non-interacting case. The dashed lines are the Nambu-GTEMPO results calculated with $\chi=250$ and $\delta t=0.05$, while the solid lines with the same colors are the corresponding ED results. (c,d) The mean errors of $G_{\uparrow\bar{\uparrow}\bullet}( t)$ (c) and $G_{\uparrow\bullet\downarrow}( t)$ (d), between Nambu-GTEMPO with $\chi=250$ and ED, as functions of the real-time step size $\delta t$. (e,f) The mean errors of $G_{\uparrow\bar{\uparrow}\bullet}( t)$ (e) and $G_{\uparrow\bullet\downarrow}( t)$ (f), between Nambu-GTEMPO with $\delta t=0.05$ and ED, as functions of the bond dimension $\chi$. We have used $\varepsilon_d = 0$ and $\beta=10$ in these simulations. 
    }
    \label{fig:toulousereal}
\end{figure}

Now we study the non-interacting case, i.e., we set $U=0$ for the impurity. We consider a continuous bath with the commonly used semi-circular BSF in Eq.(\ref{eq:semicircular}) with $D=1$.
In the non-interacting case, the bond dimension $\chi$ of the IF is no longer bounded by a constant, therefore the MPS bond truncation error would generally occur, and we will perform error analysis against $\chi$ in addition to the Trotter decomposition error. We will fix $\varepsilon_d=0$ in this study (which does not simplify our Nambu-GTEMPO calculation). For our ED calculations of this case, we have discretized the bath with a equal-distant frequency step size $\delta \omega=0.01$ and checked that our ED results have well converged against $\delta \omega$.

Similar to our study of the toy model, we first study the imaginary-time evolution in Fig.~\ref{fig:toulouseimag}, in which we set $\beta=10$. 
In Fig.~\ref{fig:toulouseimag}(a,b), we plot $\Gop_{\uparrow\uparrow}(\tau)$ and $\Gop_{\uparrow\downarrow}(\tau)$ as functions of the imaginary time $\tau$, from which we can see that our Nambu-GTEMPO results match very well with the ED results. In Fig.~\ref{fig:toulouseimag}(c,d), we plot the mean errors of the two correlation functions in Fig.~\ref{fig:toulouseimag}(a,b) between Nambu-GTEMPO and ED as functions of the imaginary-time step size $\delta\tau$, where we observe monotonically decrease of $\mathcal{E}$ as $\delta\tau$ decreases. In In Fig.~\ref{fig:toulouseimag}(e,f), we plot the mean errors of the two correlation functions in Fig.~\ref{fig:toulouseimag}(a,b) between Nambu-GTEMPO and ED as functions of $\chi$, where we can see that the mean errors decrease significantly when $\chi$ increases from $50$ to $100$, and saturates afterwards.

In Fig.~\ref{fig:toulousereal}, we further study the non-equilibrium real-time evolution for the non-interacting case, in which we set $\beta=10$. In Fig.~\ref{fig:toulousereal}(a,b), we plot the imaginary part of $G_{\uparrow\bar{\uparrow}\bullet}( t)$ (a) and the real part of $G_{\uparrow\bullet\downarrow}( t)$ as functions of the real time $t$ (the real part of $G_{\uparrow\bar{\uparrow}\bullet}( t)$ and the imaginary part of $G_{\uparrow\bullet\downarrow}( t)$ vanish in this case, and thus are not plotted).
We can see that the Nambu-GTEMPO results match very well with ED results. In Fig.~\ref{fig:toulousereal}(c,d), we plot the mean errors $\mathcal{E}$ of the two correlation functions in Fig.~\ref{fig:toulousereal}(a,b) between Nambu-GTEMPO and ED as functions of the real-time step size $\delta t$, where we observe that $\mathcal{E}$ decreases almost linearly as $\delta t$ decreases. In Fig.~\ref{fig:toulousereal}(e,f), we plot $\mathcal{E}$ of the two correlation functions in Fig.~\ref{fig:toulousereal}(a,b) between Nambu-GTEMPO and ED as functions of $\chi$ used in Nambu-GTEMPO, where we can see that $\mathcal{E}$ decrease significantly when $\chi$ increases from $50$ to $100$, and saturates afterwards, similar to Fig.~\ref{fig:toulouseimag}(e,f).

\section{Exact diagonalization algorithm in the non-interacting case}\label{app:noninteractinged}
For the non-interacting case with $U=0$, the total Hamiltonian for the SAIM becomes quadratic, which can be efficiently and exactly solved via the exact diagonalization algorithm as shown in the following.

Assuming that there are $N+1$ electron orbitals, with $1$ orbital for the impurity and $N$ orbitals for the bath, then the number of annihilation operators plus and the number of creation operators in the Hamiltonian is $L = 4(N+1)$ in total, which can be arranged into a column vector:
\begin{align}\label{eq:Cop}
\bm{\Cop}=[\aop_{\uparrow},\aop_{\downarrow},\cop_{1,\uparrow},\cop_{1,\downarrow},\cdots,\allowbreak\cop_{N,\uparrow},\cop_{N,\downarrow},\adop_{\uparrow},\adop_{\downarrow},\cdots,\allowbreak\cdop_{N,\uparrow},\cdop_{N,\downarrow}]^T.
\end{align}
With Eq.(\ref{eq:Cop}), the total Hamiltonian of the non-interacting SAIM can be written as
\begin{align}
  \Hop=\sum_{i,j=1}^{L}\Cdop_iH_{ij}\Cop_k=\bm{\Cop}^{\dag}H\bm{\Cop},
\end{align}
where the $4(N+1)\times 4(N+1)$ coefficient matrix $H$ is
\begin{align}
  H=\mqty[h&g\\g^{\dag}&0],
\end{align}
with $h$ a $2(N+1)\times2(N+1)$ coefficient matrix for the particle number-conserving part of $\Hop$, $g$ a $2(N+1)\times2(N+1)$ coefficient matrix for the non-number-conserving part. The explicit forms of $h$ and $g$ are
\begin{align}
  h=\mqty[
  \varepsilon_d & 0 & V_1 & V_2 & \cdots & V_N\\
  0 & \varepsilon_d & V_1 & V_2 & \cdots & V_N\\
  V_1 & V_1 & \varepsilon_1 &0 & \cdots & 0\\
  V_2 & V_2 & 0 & \varepsilon_2 & \cdots & \vdots\\
  \vdots & \vdots & \vdots &0 & \ddots &0\\
  V_N& V_N & 0 & \cdots &0 & \varepsilon_N\\
  ],
\end{align}
and
\begin{align}
  g=\left[\begin{array}{cccccccccc}
    0 & 0 & 0 & 0 & \cdots &\cdots & 0 & \cdots & \cdots & 0\\
    0 & 0 & 0 & 0 & \cdots &\cdots & 0 & \cdots & \cdots & 0\\
    0 & 0 & 0 & \scgap_1 & 0 & \cdots & 0 & \cdots & \cdots & 0\\
    0 & 0 & 0 & 0 & 0 & \scgap_2 &\cdots & 0 & \cdots & \vdots\\
    0 & 0 & 0 & 0 & \cdots & \ddots & \vdots & \vdots & 0 & 0\\
    0 & 0 & 0 & 0 & \cdots & \ddots & 0 & \scgap_{N-1} & 0 & 0 \\
    0 & 0 & 0 & 0 & \cdots & \cdots & 0 & 0 & 0 & \scgap_N \\
  \end{array}
  \right],
\end{align}
respectively.
Since $\Hop$ is quadratic, the equation of motion for $\Cop_j(t)$ in the Heisenberg picture is given by
\begin{align}    
  \im\dv{\Cop_j(t)}{t}=[\Cop_j(t),\Hop]=\sum_{kl}[C_j(t),H_{kl}\Cdop_{k}(t)\Cop_l(t)],
\end{align}
which gives
\begin{equation}
  \im\dv{\Cop_j(t)}{t}=\sum_{l}M_{jl}\Cop_l(t),
\end{equation}
i.e.,
\begin{align}
  \im\dv{\bm{\Cop}(t)}{t}=M\bm{\Cop}(t),
\end{align}
where the coefficient matrix $M$ is
\begin{equation}
  M=\mqty[h&g-g^{\dag}\\g^{\dag}-g&-h^{\dag}].
\end{equation}
The solution is
\begin{align}\label{eq:solution}
  \bm{\Cop}(t)=e^{-\im Mt}\bm{\Cop}.
\end{align}
Based on Eq.(\ref{eq:solution}), one could easily calculate any multi-time correlations, for example, the real-time correlation functions in Eq.(\ref{eq:gfreal})
can be obtained using
\begin{subequations}
\begin{align}
  &G^>_{\uparrow\uparrow}(t)=\frac{1}{Z}\Tr\left[(e^{-\im Mt}\bm{\Cop})_1\Cdop_1\rhoop_0 \right], \\
  &G^<_{\uparrow\uparrow}(t)=\frac{1}{Z}\Tr\left[(\bm{\Cdop}e^{\im Mt})_1\Cop_1\rhoop_0 \right], \\
  &G^>_{\uparrow\downarrow}(t)=\frac{1}{Z}\Tr\left[(e^{-\im Mt}\bm{\Cop})_1\Cdop_2\rhoop_0 \right], \\
&G^<_{\uparrow\downarrow}(t)=\frac{1}{Z}\Tr\left[(\bm{\Cdop}e^{\im Mt})_1  \rhoop_0\Cop_2 \right],  
\end{align}
\end{subequations}
where $Z=\Tr(\rhoop_0)$ and $(\bm{A})_1$ means to take the first element of the vector $\bm{A}$.

\end{document}